\documentclass[11pt,a4paper]{article}
\usepackage{jheppub}
\usepackage{mathrsfs}
\usepackage{amsfonts}
\usepackage{mathtools}
\usepackage{setspace}
\usepackage{cellspace}
\usepackage{amsmath,amssymb,bm}
\usepackage[colorlinks=true,linkcolor=blue]{hyperref}
\usepackage{xcolor}
\usepackage{epsfig}
\usepackage{slashed}
\usepackage{subcaption}
\usepackage{hhline,multirow,tabularx}  
\usepackage{dcolumn}    
\usepackage{url}        
\usepackage{braket}     
\usepackage{pifont}
\newcommand{\aem}{\alpha_{\rm{EM}}}

\usepackage{physics}

\title{Small-$x$ gluon GPD constrained from deeply virtual $J/\psi$ production and gluon PDF through universal-moment parameterization}

\author[a]{Yuxun~Guo}
\author[b]{, Xiangdong~Ji} 
\author[c,d]{, M.~Gabriel~Santiago}
\author[b]{, Jinghong~Yang}
\author[e]{and Hao-Cheng~Zhang}

\affiliation[a]{Nuclear Science Division, Lawrence Berkeley National Laboratory,\\ Berkeley, CA 94720, USA}
\affiliation[b]{Maryland Center for Fundamental Physics, Department of Physics, University of Maryland,\\ 4296 Stadium Dr., College Park, MD 20742, USA}
\affiliation[c]{Center for Nuclear Femtography, SURA,\\ 1201 New York Ave. NW, Washington, DC 20005, USA}
\affiliation[d]{Department of Physics, Old Dominion University,\\ Norfolk, VA 23606, USA}
\affiliation[e]{Taishan College, Shandong University,\\ Jinan, Shandong, 250100, China}

\emailAdd{yuxunguo@lbl.gov}
\emailAdd{xji@umd.edu}
\emailAdd{gsantiago@sura.org}
\emailAdd{yangjh@umd.edu}
\emailAdd{hczhang2003@gmail.com}

\abstract{We phenomenologically constrain the small-$x$ and small-$\xi$ gluon generalized parton distributions (GPDs) with the deeply virtual $J/\psi$ production (DV$J/\psi$P) in the framework of GPDs through universal moment parameterization (GUMP). We use a hybrid cross-section formula combining collinear factorization to the next-to-leading order (NLO) accuracy of the strong coupling $\alpha_s$, with corrections from non-relativistic QCD to account for the power corrections due to the heavy $J/\psi$ mass. We reach reasonable fit to the measured differential cross-sections of DV$J/\psi$P by H1 at Hadron-Electron Ring Accelerator (HERA) as well as forward gluon PDFs from JAM22 global analysis. We find that both NLO and non-relativistic corrections are significant for heavy vector meson productions. Of course, the gluon GPD we obtain still contain considerable freedom in need of inputs from other constraints, particularly in 
the distribution-amplitude-like region.}

\keywords{Generalized parton distributions; Deeply virtual $J/\psi$ production; GUMP;}
\date{\today}
\begin{document}
\maketitle


\section{Introduction}

The multidimensional structure of hadrons, as encoded by their quark and gluon degrees of freedom in quantum chromodynamics (QCD), remains a major open problem in nuclear physics. Several powerful tools for addressing this very problem have been developed in the past few decades, one of which is the generalized parton distributions (GPDs) \cite{Muller:1994ses, Ji:1996ek, Ji:1998pc}. GPDs leverage nonzero momentum transfer to a hadron to go beyond the forward parton distribution functions (PDFs) that carry the one-dimensional, longitudinal structure, and instead probe the three-dimensional structure in the impact parameter space~\cite{Burkardt:2000za,Burkardt:2002hr}. Thus, GPDs are investigated for the tomography of hadrons~\cite{Burkardt:2000za,Burkardt:2002hr,Ji:2003ak, Belitsky:2003nz} as well as various properties of the bulk hadron state, such as the distributions of spin and mass in the quark and gluon fields comprising the bound state~\cite{Ji:1994av, Ji:1996ek, Polyakov:2002yz}. These functions are related to exclusive production processes like deeply virtual Compton scattering (DVCS) and deeply virtual meson production (DVMP) through factorization theorems~\cite{Ji:1996nm,Radyushkin:1996ru,Collins:1996fb}. However, the amplitudes in these processes are given by convolutions involving GPDs together with other quantities, which essentially eliminates one of their three variables from being directly probed. This poses an inverse problem for the deconvolution of these amplitudes to extract GPDs, as illustrated in the context of shadow GPDs recently~\cite{Bertone:2021yyz}.

Accordingly, massive efforts have been devoted to searching for complimentary inputs. Recent measurements of time-like Compton scattering~\cite{CLAS:2021lky} and near-threshold $J/\psi$ productions~\cite{GlueX:2019mkq,Duran:2022xag,GlueX:2023pev} provide some new insights to access the GPDs through exclusive photo-productions~\cite{Berger:2001xd,Ivanov:2004vd,Guo:2021ibg,Guo:2023pqw,Guo:2023qgu}. Moreover, exclusive productions of multiple particles that typically have better sensitivity to the missing variable have been proposed~\cite{Belitsky:2002tf,Guidal:2002kt,Pedrak:2017cpp,Duplancic:2023kwe}, and a new factorization theorem for single diffractive hard exclusive processes has been developed recently~\cite{Qiu:2022bpq,Qiu:2022pla,Qiu:2023mrm,Nabeebaccus:2023rzr,Qiu:2024mny}. Probing the nucleon structures with these exclusive processes will be one of the crucial tasks of the future Electron-Ion Collider (EIC)~\cite{AbdulKhalek:2021gbh}. On the other hand, the lattice QCD offers a different perspective to approach the nuclear structure in Euclidean space from first principle. The simulations of the nucleon form factors (FFs) on lattice have been significantly improved just in the past few years and show exciting results~\cite{Alexandrou:2021jok,Hasan:2017wwt,Shintani:2018ozy,Jang:2018djx,Bhattacharya:2023ays,Bhattacharya:2023jsc,Pefkou:2021fni,Hackett:2023rif}. Moreover, developments like the large momentum effective theory (LaMET) have enabled the lattice QCD to directly calculate the parton distributions including GPDs~\cite{Ji:2013dva,Ji:2020ect}, with which many pioneering works have been done for the GPDs~\cite{Alexandrou:2020zbe,Lin:2021brq,Bhattacharya:2022aob}. Therefore, we will need a global analysis program of GPDs, that combines the comprehensive present and future inputs on GPDs including both experiment measurements and lattice simulations and perhaps overcome the inverse problem once enough constraints are gathered. 

In the previous works of the authors~\cite{Guo:2022upw,Guo:2023ahv}, the GPDs through Universal Moment Parameterization (GUMP) framework was put forward for performing such a global analysis by making use of the conformal moment expansion of GPDs \cite{Mueller:2005ed, Kumericki:2009uq}. A preliminary global analysis was performed by combining input from globally extracted PDFs~\cite{Cocuzza:2022jye} and FFs~\cite{Ye:2017gyb}, lattice calculations~\cite{Alexandrou:2021jok,Alexandrou:2020zbe}, and DVCS data from Jefferson Lab (JLab) and the Hadron-Electron Ring Accelerator (HERA)~\cite{CLAS:2018ddh,CLAS:2021gwi,Georges:2017xjy,JeffersonLabHallA:2022pnx,H1:2009wnw} to constrain the up and down quark GPDs, both in the valence and sea quark regimes at leading order (LO) in the perturbative expansion in $\alpha_s$~\cite{Guo:2023ahv}. The gluon GPDs were only constrained in the forward (PDF) limit, as the DVCS cross-sections are not sensitive to them until at next-to-leading order (NLO) or through the evolution of the sea quarks. In this paper, we extend the analysis by adding the off-forward gluon GPDs together with the deeply virtual $J/\psi$ production (DV$J/\psi$P) measurements at HERA by the H1 collaboration~\cite{H1:2005dtp}. Assuming the intrinsic quark content of the target is negligible for heavy quarks like the charm, the heavy quarkonium productions will be mostly sensitive to the gluon GPDs, especially in the gluon-dominant high $Q^2$ and small-$x_B$ kinematics for the HERA measurements. Due to this unique sensitivity to the gluonic structure in the nucleon, there have been extensive studies on the exclusive productions of the $J/\psi$ in various frameworks~\cite{Ryskin:1992ui,Brodsky:1994kf,Frankfurt:1997fj,Frankfurt:2000ez,Ivanov:2004vd,Kowalski:2006hc,Chen:2019uit, Koempel:2011rc,Koempel:2015xol,Flett:2021ghh,Mantysaari:2021ryb,Mantysaari:2022kdm,Eskola:2022vpi,Goloskokov:2024egn,Flett:2024htj}. 

The exclusive photo-production of the $J/\psi$, corresponding to the $Q\to 0$ limit of lepto-production, has been calculated to NLO in~\cite{Ivanov:2004vd} in the framework of non-relativistic QCD (NRQCD), where the heavy meson mass serves as the hard scale for the factorization. It has been later extended to electro-production~\cite{Chen:2019uit, Flett:2021ghh}, where the NLO effects are shown to be significant and could strongly cancel the LO contributions with some simple GPD model~\cite{Chen:2019uit}.
However, an analysis strictly within the GPD framework with corresponding NLO evolution has not yet been established. On the other hand, there have been many developments in the analysis of light meson productions in the collinear factorization framework of DVMP, for which the DVMP of light meson as well as DVCS have been studied up to NLO including GPD evolution~\cite{Goloskokov:2006hr,Meskauskas:2011aa,Lautenschlager:2013uya,Muller:2013jur,Duplancic:2016bge,Cuic:2023mki}. This process has also been studied in the small-$x$ color-glass-condensate (CGC) framework \cite{Kovchegov:1999ji,Kovner:2001vi,Hentschinski:2005er,Kovner:2006ge,Hatta:2006hs,Kowalski:2006hc,Rezaeian:2012ji} and recently even at NLO~\cite{Mantysaari:2021ryb,Mantysaari:2022kdm}. However, an approximate matching between the CGC and the collinear framework has been established for the DVCS~\cite{Hatta:2017cte}, but not for the DV$J/\psi$P yet. These developments provide us the opportunity to consider the heavy vector meson production in a strict GPD framework to NLO including evolution, of which the theoretical setup will be introduced in the next section. 

The structure of the rest of this paper is as follows, in Sec.~\ref{sec_glue}, we introduce the theoretical setup used to study the gluon GPDs with the DV$J/\psi$P cross-section, paying special attention to the various difficulties which have arisen in past efforts to extract the gluon GPD or PDF from this process. Next, in Sec.~\ref{sec_jp} we specify how we implement constraints from DV$J/\psi$P in the GUMP framework and present the results of our new fit. First, we review the results of our previous global analysis of up and down quarks \cite{Guo:2023ahv}, then we clarify the parameters and formulas used for the modelling of the GPDs and the calculations of the cross-section when performing our fit to the data. We then present the results of our fits and compare them to the input DV$J/\psi$P data and PDFs. Finally, in Sec.~\ref{sec_con} we summarize our results and discuss directions for further work.


\section{Exclusive $J/\psi$ electro-production for the gluonic structure}
\label{sec_glue}
\begin{figure}[t]
\centering
\includegraphics[width= 0.9 \linewidth]{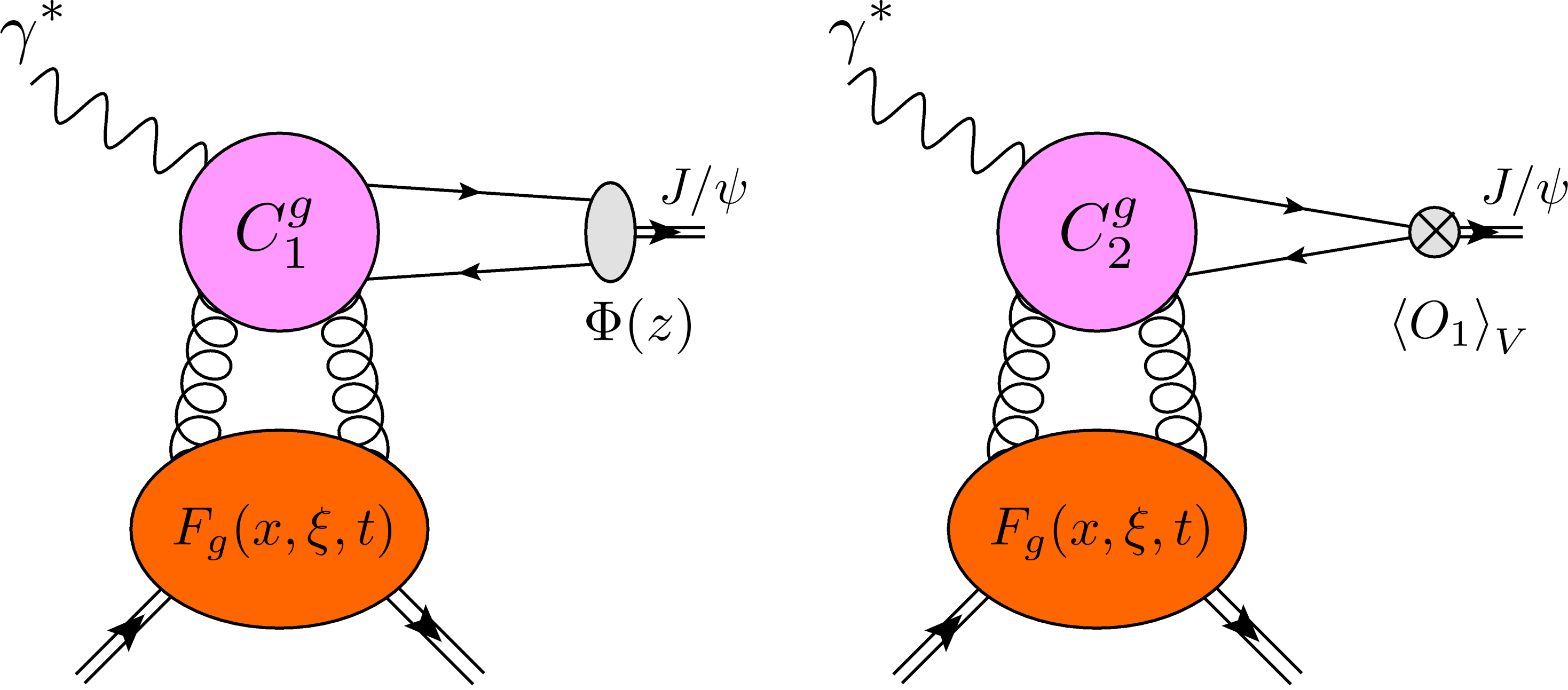}
\caption{Illustrations of the factorized amplitudes for in the collinear factorization framework (left) and the NRQCD framework (right). The DA $\Phi(z)$ in the collinear framework will be replaced by a local NRQCD matrix element $\left<O_1\right>_V$ in the NRQCD framework, with different hard scattering coefficients $C_1^g(x,\xi,z)$ and $C_2^{g}(x,\xi,M_V,Q)$ to be matched.}
\label{FIG:facillu}
\end{figure}

In this section, we introduce a factorization formula for the electro-production of heavy vector meson, which will be used to constrain the gluon GPD in this paper. In the deeply virtual, i.e., $Q\to \infty$ limit where $Q^2\equiv -q^2$ is the virtuality of the photon with momentum $q$, a collinear factorization theorem has been proven at leading-twist for longitudinally polarized virtual photon~\cite{Collins:1996fb,Radyushkin:1996ru}. Meanwhile, for exclusive heavy meson productions, the non-relativistic QCD (NRQCD) factorization has been established for both the photo- and electro-production to NLO in the limit that the meson mass $M_V\to\infty$~\cite{Ivanov:2004vd,Chen:2019uit,Flett:2021ghh}. A diagrammatic illustration of the two frameworks is shown in Fig. \ref{FIG:facillu}. Nevertheless, the two frameworks have been explored in rather different context, focusing on light-meson productions~\cite{Muller:2013jur,Cuic:2023mki} and heavy-meson productions~\cite{Ivanov:2004vd,Chen:2019uit,Flett:2021ghh}, respectively, whereas their connection beyond the LO has not yet been established.

\begin{table}[h] 
    \def\arraystretch{1.25}
    \centering
    \begin{tabular}{>{\centering\arraybackslash} m{6cm}|>{\centering\arraybackslash} m{3cm} >{\centering\arraybackslash} m{3cm}}
    \hline\hline
    Features & DVMP & NRQCD  \\ \hline \hline
    Light meson production & {\color{red}\ding{52}} & - \\ \hline
    Simple moment space expressions & {\color{red}\ding{52}} & \scalebox{1.25}{\textbf{?}} \\ \hline
    Meson mass correction & - & {\color{red}\ding{52}}\\ \hline
    Transverse photon cross-section & - & {\color{red}\ding{52}} \\ \hline
    Fixed-order results up to NLO & {\color{red}\ding{52}} & {\color{red}\ding{52}}\\
    \hline \hline
    \end{tabular}
    \caption{Comparison of the different features of the collinear factorization and NRQCD factorization framework.}
    \label{tab:frameworkcomp}
\end{table}

While each framework has been shown successful for its own application, the lack of proper matching between would cause issues for a global analysis of GPDs. On the one hand, a combined analysis of the light and heavy meson productions would allow enhanced sensitivities to GPDs, especially to disentangle their flavor structures. On the other hand, each framework has its limitations, as summarized in Table \ref{tab:frameworkcomp}, which can be compensated by a hybrid framework.
Ideally, a hybrid factorization framework that perturbatively matches the collinear and NRQCD factorization frameworks when $Q^2$ and $M_V^2$ are both large, and reduces to them under different limits respectively could overcome these issues. However, as we will show below that while the matching is trivial at LO, it gets much more complicated at NLO which requires more detailed theoretical studies. In this work, we will consider an approximate hybrid framework based on the LO matching while including the NLO corrections from the collinear framework, whereas refined studies on the NLO matching will be left to future work. This is the best approximation we can come up with based on the state-of-art theoretical work~~\cite{Ivanov:2004vd,Muller:2013jur,Cuic:2023mki,Chen:2019uit,Flett:2021ghh}.

In the following of this section, we will briefly review the two factorization frameworks and introduce the hybrid framework of this work. 

In the collinear factorization of DVMP, the differential cross-sections of DVMP for an unpolarized proton target can be written in the small-$x_B$ limit as~\cite{Cuic:2023mki}:
\begin{equation}
    \frac{\text{d}\sigma^{\rm{DVMP}}_{\gamma^*_L p\to J/\psi\; p}}{\text{d}t}=4\pi^2 \aem \frac{x_B^2}{Q^4}\left(\left|\mathcal{H}_{\rm{DVMP}}\right|^2-\frac{t}{4 M_N^2} \left|\mathcal{E}_{\rm{DVMP}}\right|^2\right) \ ,
\end{equation}
where $\aem$ is the fine structure constant and $\mathcal{H}_{\rm{DVMP}}$ and $\mathcal{E}_{\rm{DVMP}}$ are the so-called transition form factors (TFFs) that correspond to the $H$ and $E$ GPDs defined below. Note that the lepto-production cross-section has been rewritten into a virtual photo-production cross-section by removing the virtual photon flux of the lepton. Thus, the total virtual photo-production cross-section can be written as the sum of the longitudinal- and the transverse-photon contributions:
\begin{equation}
    \frac{\text{d}\sigma_{\gamma^* p\to J/\psi\; p}}{\text{d}t}  = \frac{\text{d}\sigma_{\gamma^*_{T} p\to J/\psi\; p}}{\text{d}t} + \varepsilon \frac{\text{d}\sigma_{\gamma^*_{L} p\to J/\psi\; p}}{\text{d}t}
\end{equation}
where $\varepsilon$ is the ratio of longitudinal and transverse photon flux that can be determined with the kinematics of the lepton beam. Moreover, since the factorization theorem is proven at leading twist with longitudinally polarized photon, only the longitudinal $\text{d}\sigma_{\gamma^*_{L} p\to J/\psi\; p}$ can be calculated in this framework.

The TFFs $\mathcal{H}_{\rm{DVMP}}$ (and $\mathcal{E}_{\rm{DVMP}}$ similarly) can then be written in the factorized form as~\cite{Muller:2013jur,Cuic:2023mki}
\begin{equation}
\label{eq:Advmp}
    \mathcal{H}^g_{\rm{DVMP}} = \frac{ e_q f_{V} }{N_C Q} \int_0^{1}\text{d}z\int\limits_{-1}^1 \dd{x} C^g_1 (x, \xi,z) H^g (x, \xi, t) \Phi^V(z)\ ,
\end{equation}
as illustrated in Fig. \ref{FIG:facillu}. In the above expression, $f_{V}$ is the meson decay constant, $V$ denotes the vector meson: $V=\rho,\omega,\phi,J/\psi,\cdots$\footnote{In this work we consider the $J/\psi$ production which is vector meson. The factorization formula applies to other mesons as well, with slight changes in the formula and GPDs involved in the TFFs~\cite{Frankfurt:1999fp,Muller:2013jur,Kroll:2019wug,Cuic:2023mki}, which will not be discussed here.}, $N_c=3$ is the number of color, and $e_q$ is the quark charge in the unit of proton charge $e$, i.e., $2/3$ for charm and $-1/3$ for strange quark. All the scale dependence will be suppressed in this section for simplicity. The constant $f_{V}$ can be determined according to physical observables such as the leptonic decay width as~\cite{Ball:2006eu},
\begin{align}
\label{eq:lepdecaywidth}
    \Gamma \left[ V \rightarrow e^+ e^- \right]  = \frac{4\pi e_q^2 \aem^2 f^2_{V}}{3 M_V}\ ,
\end{align}
where $M_V$ is the vector meson mass. Taking from the particle data group~\cite{ParticleDataGroup:2020ssz} that $\Gamma[J/\psi \rightarrow e^+ e^-]=5.53\pm0.10$ keV, one gets $f_{J/\psi}\approx 0.41$ GeV.

Inside the integral for the TFFs, there are three components: the meson distribution amplitude (DA) $\Phi^V(z)$ 
is defined by~\cite{Ball:1998sk,Ball:1998ff}:
\begin{align}
    M_V f_V \frac{n\cdot \varepsilon^V}{n\cdot P_V} \Phi_q^V(z)= \frac{1}{P_V^+} \int^\infty_{-\infty} \frac{\text{d}\lambda}{2\pi} e^{i(2z-1) \lambda/2} \left<0\left|\bar \psi_q\left(-\frac{\lambda n}{2}\right)\gamma^+\psi_q\left(\frac{\lambda n}{2}\right) \right|V_L(P_V)\right>\ ,
\end{align}
normalized according to $\int^1_0 \text{d}z \Phi(z) =1$, where $n$ is the light-cone vector, conjugating to the meson momentum $P_V$. At leading twist, the vector meson is longitudinally polarized and so is the corresponding DA. The leading-twist gluon GPDs $F^g(x,\xi,t)$ are defined by~\cite{Ji:1996ek,Diehl:2003ny},
\begin{align}
   F_g(x,\xi,t) \equiv &\frac{1}{(\bar P^+)^2} \int \frac{\text{d}\lambda}{2\pi} e^{i\lambda x}\left<P'\left|F^{+\mu}_{\;\;\;\;\;}\left(-\frac{\lambda n}{2}\right)F^{+}_{\;\;\mu}\left(\frac{\lambda  n}{2}\right) \right|P\right>\ ,
\end{align}
where $P$ and $P'$ are the momenta of the initial and final nucleon with $\bar P \equiv (P+P')/2$ and $\Delta\equiv P'-P$, $x$ is the average parton longitudinal momentum, $\xi\equiv -n\cdot \Delta/(2n\cdot \bar P)$ is the skewness parameter and $t\equiv \Delta^2$ is the squared momentum transfer. They can be further parameterized in terms of scalar functions as~\cite{Ji:1996ek}
\begin{align}
    F_{g}(x,\xi,t)=\frac{1}{2\bar P^+}\bar u(P')\left[H_g(x,\xi,t) \gamma^+ +E_g(x,\xi,t)\frac{i\sigma^{+\alpha}\Delta_{\alpha}}{2 M_N}\right]u(P) \ ,
\end{align}
where $H_g(x,\xi,t)$ and $E_g(x,\xi,t)$ are the well-known leading-twist gluon GPDs~\cite{Ji:1996ek}; the hard scattering coefficient $C_1^g(x,\xi,z)$ which can be perturbatively calculated. There will also be contributions from quark GPDs $F_q(x,\xi,t)$ with their corresponding Wilson coefficient $C_{1}^q(x,\xi,z)$ starting at NLO, whose expressions are not explicitly presented here.

This framework has been applied to analyze the measurement of DVMP production of light meson at HERA at NLO, which shows great success~\cite{Muller:2013jur,Cuic:2023mki}. Nevertheless, it is noteworthy that the vector meson mass neglected in the DVMP formula can introduce sizable corrections in the case of $J/\psi$. Even for the highest $Q^2$ bin in HERA data \cite{ZEUS:2004yeh,H1:2005dtp}, one has $M_{J/\psi}^2 / \left< Q^2 \right> \approx 1/2$. These corrections in the form of $M_{J/\psi}^2 / Q^2$ will be higher-twist effects, which lie outside the approximations of the collinear factorization theorem \cite{Collins:1996fb}. 

On the other hand, these higher-twist terms, including the contribution from transversely polarized virtual photon, can be calculated within the NRQCD framework, where the heavy meson mass $M_{J/\psi}$ is treated as the hard scale and the formation of a non-relativistic meson bound state will be factorized into a local NRQCD matrix element. This has been calculated up to the NLO for the photo-production~\cite{Ivanov:2004vd} as well as the lepto-production~\cite{Chen:2019uit,Flett:2021ghh}. In this framework, the corresponding differential cross-section in the small $x_B$ limit reads~\cite{Ivanov:2004vd,Koempel:2011rc}
\begin{equation}
    \frac{\text{d}\sigma^{\rm{NRQCD}}_{\gamma^*_L p \to J/\psi\;p}}{\text{d}t}=4\pi^2 \aem \frac{x_B^2}{\left(Q^2+M_{J/\psi}^2\right)^2}\left(\left|\mathcal{H}_{\rm{NRQCD}}\right|^2-\frac{t}{4 M_N^2} \left|\mathcal{E}_{\rm{NRQCD}}\right|^2\right) \ .
\end{equation}
which formally reduce to the DVMP cross-section in the limit $Q\gg M_{J/\psi}$, though the amplitudes will be differently. \footnote{Meanwhile, we also note that the application of NRQCD factorization is not limited to just the high-energy productions of heavy mesons, but also lower-energy near-threshold photo-productions~\cite{Guo:2021ibg}, since the heavy meson mass serves as the hard scales.} In the NRQCD framework, the corresponding amplitudes can be written as~\cite{Ivanov:2004vd},
\begin{align} \label{amp_fac}
    \mathcal{H}^g_{\rm{NRQCD}} = \frac{  e_q}{N_C } \left(\frac{\left<O_1\right>_V}{m_q}\right)^{1/2} \frac{Q}{Q^2+M_{V}^2} \int\limits_{-1}^1 \dd{x} C^g_2 \left(x, \xi, M_V,Q\right) H^g (x, \xi, t) \ ,
\end{align}
where $m_q$ is the corresponding quark mass, which can be taken to be $M_V/2$ for heavy quarkonia like $J/\psi$ or $\Upsilon$, omitting the higher-order non-relativistic corrections suppressed by the heavy quark velocity $v \sim \alpha_s(M_V)$.

The NRQCD matrix element $\left<O_1\right>_V$ is defined as $\left<O_1\right>_V\equiv \left<J/\psi\right|\hat O_1(^3S_1)\left|J/\psi\right>$, where the local NRQCD operator $\hat O_1(^3S_1)$ can be written as $\psi^\dagger \boldsymbol{\sigma} \chi \cdot \chi^\dagger \boldsymbol{\sigma} \psi$ in terms of the heavy quark field $\psi$ and antiquark field $\chi$.
On the other hand, this NRQCD matrix element $\left<O_1\right>_V$ can also be related to the meson decay width~\cite{Bodwin:1994jh}:
\begin{align}
\label{eq:lepdecaywidth2}
    \Gamma \left[ V \rightarrow e^+ e^- \right]  = \frac{2\pi e_q^2 \aem^2 \left<O_1\right>_V}{3 m_q^2}\ .
\end{align}
Comparing to eq. (\ref{eq:lepdecaywidth}), one has the simple relation $\left<O_1\right>_V=m_q f_V^2$, with which we have
\begin{align} \label{amp_fac2}
    \mathcal{H}^g_{\rm{NRQCD}} = \frac{  e_q f_V}{N_C } \frac{Q}{Q^2+M_{V}^2} \int\limits_{-1}^1 \dd{x} C^g_2 \left(x, \xi, M_V,Q\right) H^g (x, \xi, t) \ ,
\end{align}
that resembles the DVMP one in eq. (\ref{eq:Advmp}). However, in the case of NRQCD, there exist additional non-relativistic expansion in powers of $v \sim \alpha_s(M_V)$, of which the corresponding corrections can be calculated as well~\cite{Hoodbhoy:1996zg}.

The matching between the two amplitudes or TFFs in the two frameworks is rather obvious at LO, for which the Wilson coefficients at LO read respectively,
\begin{equation}
    C^{g,(\rm{LO})}_1 (x, \xi, z)= \frac{1}{1-z} \frac{\alpha_s}{\xi\left(\xi-x-i\epsilon\right)}\ ,
\end{equation}
and 
\begin{equation}
    C^{g,(\rm{LO})}_2 (x, \xi)= \frac{2\alpha_s}{\xi\left(\xi-x-i\epsilon\right)}\ ,
\end{equation}
that does not depend on the meson mass $M_V$ or $Q$. They are almost identical except for the different prefactors, which can be matched after the $\text{d}z$ integral if a non-relativistic approximation of the DA $\Phi(z)$ is made --- in NRQCD, the relative velocity $v$ of the two quark in the meson is suppressed by the strong coupling constant $v\sim\alpha_S$. Thus, in the LO picture, each of them shares half of the total meson momentum with $z=1/2$ and we have: $\Phi_{\rm{NR}}(z) = \delta\left(z-1/2\right)$ and thus $\int \text{d} z\;\Phi_{\rm{NR}}(z)/(1-z) = 2$. This corresponds to the extra factor of $2$ in the $C_2^{g,(\rm{LO})} (x, \xi)$. Thus, the amplitudes in the two frameworks become identical in the limit $Q\gg M_V$.

While this can be trivially done at LO as shown above, further complexity arises at higher order, which is found necessary in our analysis of the DV$J/\psi$P data --- the attempt to fit the data fails with only the LO hard scattering amplitude and QCD scale evolution, indicating significant NLO corrections.  While the NLO corrections have been calculated in the NRQCD framework~\cite{Chen:2019uit,Flett:2021ghh} as well as the DVMP framework~\cite{Duplancic:2016bge,Cuic:2023mki}, there does not exist a smooth transition between them. In NRQCD, the loop integral will be naturally regulated by the mass quark propagators, leading to log terms in the form of $\log(m_c^2/\mu_F^2)$ with $\mu_F$ the factorization scale~\cite{Chen:2019uit,Flett:2021ghh}, of which the limit $m_c\to 0$ does not exist, whereas in the collinear framework, there are no such dimensional quantities and thus the logs will be in the form of $\log(1/\xi)$~\cite{Muller:2013jur}, besides the ones that can be factorized into the meson DA. These single-log terms can also combine to form double logs. Consequently, the matching of the two hard scattering coefficients becomes extremely non-trivial. Therefore, we consider a hybrid framework that utilizes the DVMP framework at NLO but only match the non-relativistic corrections at LO, whereas a refine study comparing and matching the two framework beyond LO will be left to future work.

To do so, we consider the DVMP amplitude with the replacement $1/Q\to Q/(Q^2+M_{J/\psi}^2)$:
\begin{align}
    \mathcal{H}^g_{\rm{Hyb.}} = \frac{ e_q f_{J/\psi} }{N_C} \frac{Q}{Q^2+M_{J/\psi}^2}\int_0^{1}\text{d}z \int\limits_{-1}^1 \dd{x} C_1^g (x, \xi,z) H^g (x, \xi, t) \Phi(z) \ ,
\end{align}
and also replace the kinematic factor in the cross-section, 
\begin{equation}
\label{eq:xsechyb}
    \frac{\text{d}\sigma^{\rm{Hyb.}}_L}{\text{d}t}=4\pi^2 \aem \frac{x_B^2}{\left(Q^2+M_{J/\psi}^2\right)^2}\left(\left|\mathcal{H}_{\rm{Hyb.}}\right|^2-\frac{t}{4 M_N^2} \left|\mathcal{E}_{\rm{Hyb.}}\right|^2\right) \ ,
\end{equation}
whereas the Wilson coefficients and $J/\psi$ DA $\Phi(z)$ from the DVMP framework will still be used. This hybrid framework allows us to not only take the leading mass corrections from the heavy $J/\psi$ mass into account in the NRQCD framework as well as the large logarithms $\log(1/\xi)$ at NLO for $\xi$ as low as $10^{-4}$, but also to extend to light vector meson production for future analyses, which cannot be achieved otherwise with the collinear or NRQCD framework alone. Still, while the replacement gives the correct LO matching between the two frameworks as shown above, the major systematics are from the mismatch of the NLO amplitudes that include both the quark and gluon contributions, which will be studied in a separate future work.

Another advantage of such a hybrid framework lies in the photon polarizations which enter the amplitude. For the collinear case, the leading-twist contribution comes from the longitudinal photon polarization, and this is the only contribution for which the factorization theorem is proven for light vector meson production. Thus, one needs to measure the ratio of the cross-section for longitudinally polarized photons to that for transversely polarized photons $R = \dd \sigma_L / \dd \sigma_T$ and use this to relate the factorized portion of the cross-section $\dd \sigma_L$ to the total cross-section through
\begin{align} \label{r_rat_xsec}
    \frac{\dd \sigma^{\rm{Hyb.}}_{tot}}{\dd t} = \frac{\dd \sigma^{\rm{Hyb.}}_L}{\dd t} \left( \varepsilon + \frac{1}{R} \right)\ ,
\end{align}
with the longitudinal cross-sections given by eq. (\ref{eq:xsechyb}). In the NRQCD treatment, the hard part of the amplitudes, including the one for transverse photon polarization, can be constructed perturbatively, which also predicts a longitudinal-transverse ratio $R = Q^2 / M_{J/\psi}^2$. This eliminates the need for experimental extraction of $R$, and thus removes one of the largest sources of uncertainty in the data~\cite{ZEUS:2004yeh,H1:2005dtp}. This prediction for $R$ also connects the amplitude smoothly to the photoproduction limit $Q^2 \rightarrow 0$, which could allow for the inclusion of much more $J/\psi$ photo-production data in future work. The NRQCD framework has also been used to study the near-threshold $J/\psi$ production \cite{Guo:2023pqw,Guo:2023qgu}. These processes can provide constraints on GPDs in very different regions of $(\xi,t)$ phase space and thus enhance the scope of a global analysis.

\section{DV$J/\psi$P with GPDs in the GUMP framework}
\label{sec_jp}

In this section, we discuss the implementation of $J/\psi$ electroproduction data as a constraint on gluon GPDs within the GUMP framework. We start by reviewing the ingredients of the GUMP framework and the previous forward \cite{Guo:2022upw} and off-forward analyses \cite{Guo:2023ahv}, where the $u$ and $d$ quark valence and sea GPDs were constrained by a combination of experimental data and lattice computations. Then we will discuss how we include $J/\psi$ data from H1 \cite{H1:2005dtp} as a constraint on the off-forward parameters entering the gluon GPDs. Finally, we will discuss the results of the fit and the extracted gluon GPDs.


\subsection{Constraints from previous work}
\label{subsec_prev}


The analysis in \cite{Guo:2023ahv} included conformal moment space models for the valence and sea quark GPDs for the $u$ and $d$ quarks, as well a model for the semi-forward limit of the gluon GPDs ($t$-dependent PDFs). In these models, the GPD conformal moments are directly parameterized, similarly to the KM models used in \cite{Kumericki:2007sa,Kumericki:2009uq,Meskauskas:2011aa,Lautenschlager:2013uya,Cuic:2020iwt,Cuic:2023mki}. The details of the conformal moment parameterization as used in the GUMP framework thus far are given in \cite{Guo:2022upw,Guo:2023ahv}, we will briefly review them here. A GPD $F^i (x,\xi, t)$ for parton species $i$ can be expressed as a series expansion in conformal moments $\mathcal{F}^i_j (\xi,t)$ as
\begin{align}
\label{eq:conformalsum}
  F^i (x,\xi,t) = \sum_{j=0}^{\infty} (-1)^j p^i_j(x,\xi) \mathcal {F}^i_{j}(\xi,t)\ ,
\end{align}
with $p^i_j(x,\xi)$ the conformal wave functions, which can be expressed in terms of Gegenbauer polynomials \cite{Mueller:2005ed, Kumericki:2009uq}. Each individual moment is ensured to be written in terms of polynomials in $\xi$ by the polynomiality condition for GPDs \cite{Ji:1998pc}:
\begin{align} \label{mom_pol}
\mathcal{F}^i_j(\xi,t)=\sum_{k=0,\rm{ even}}^{j+1} \xi^{k} \mathcal{F}^i_{j,k}(t) \ ,
\end{align}
and for the GUMP framework we take the generalized form factors $\mathcal{F}^i_{j,k}(t)$ to be given by
\begin{align} \label{mom_ans}
    \mathcal{F}^i_{j,k}(t) = \sum_{l=1}^{l_{\rm{max}}}N^i_{l,k} B(j+1-\alpha_{l,k},1+\beta_{l,k})\frac{j+1-k-\alpha_{l,k}}{j+1-k-\alpha_{l,k}(t)} f(t) .
\end{align}
Here we have overall normalization constants $N^i_{l,k}$, Euler beta functions coming from the forward PDF $x$-dependence as $x^{-\alpha_{l,k}}(1-x)^{\beta_{l,k}}$, the Regge trajectory $\alpha_{l,k}(t) = \alpha_{l,k} + \alpha'^{i}_{l,k} t$, and the `residual' term $f(t) = e^{b^i_{l,k}t}$ which gives the exponential falloff in $t$ which is observed at small-$x$. For simplicity, as well to keep the number of free parameters from over fitting the data we use for the analysis, we take a single term in the ansatz \eqref{mom_ans}, setting $l_{\rm{max}} = 1$.

Furthermore, we impose several extra constraints to reduce the number of free parameters: first the Regge intercepts of the sea quarks are taken to be equal, $\alpha'^{\bar{u}} = \alpha'^{\bar{d}} $, as well as their `residual' terms, $b^{\bar{u}} = b^{\bar{d}}$. These quark parameters will be fixed from the previous DVCS analysis \cite{Guo:2023ahv}, whereas the corresponding gluon parameters, $\alpha'^g$ and $b^g$, will be determined from experimental data. Second, we take the higher-order terms in the $\xi$ polynomial for the moments \eqref{mom_pol} to be proportional to the leading $k=0$ term so that
\begin{align}
    \mathcal{F}^i_j(\xi,t)=\sum_{k=0,\rm{ even}}^{j+1} \xi^{k} R^i_{\xi^k} \mathcal{F}^i_{j,0}(t) ,
\end{align}
with the sum being truncated at $k = 2$ in \cite{Guo:2023ahv}, taking advantage of the high-energy limit where $\xi$ is relatively small. Here, we extend to include a $k=4$ term to allow extra flexibility in modeling the $\xi$-dependence for the gluon. Finally, some GPD species are not as well constrained by currently available data, and so we took them to be proportional to more well constrained GPD species, for example setting the sea quark $E$ GPDs to be proportional to the sea quark $H$ GPDs, $E^{\bar{q}} = R^{E}_{\rm{sea}} H^{\bar{q}}$. The same applies to the gluon GPD: $E^{g} = R^{E}_{\rm{sea}} H^{g}$ in the previous fit. As the sensitivity to the gluon $E_g$ GPDs is suppressed in the small-$t$ region as shown in eq. (\ref{eq:xsechyb}), the $R^{E}_{\rm{sea}}$ will be fixed from previous results too, modifying which appears to have small effects on the DV$J/\psi$P cross-sections.

\begin{table}[h] 
    \def\arraystretch{1.5}
    \centering
    \begin{tabular}{>{\centering\arraybackslash} m{1.8cm}|>{\centering\arraybackslash} p{2.5cm} |>{\centering\arraybackslash} p{2,5cm}|>{\centering\arraybackslash} p{2.5cm}|>{\centering\arraybackslash} p{2.5cm}}
    \hline\hline
    Parameters & $H$ & $E$ & $\widetilde{H}$  & $\widetilde{E}$ \\ \hline \hline
    \multirow{2}{*}{$u_V$ }&$N^{H}_{u_V}$,$\alpha^{H}_{u_V}$,$\beta^{H}_{u_V}$, & $N^{E}_{u_V}$,$\alpha^{E}_{u_V}$,$\beta^{E}_{u_V}$, & $N^{\tilde{H}}_{u_V}$,$\alpha^{\tilde{H}}_{u_V}$,$\beta^{\tilde{H}}_{u_V}$, & $N^{\tilde{E}}_{u_V}$,$\alpha^{\tilde{E}}_{u_V}$,$\beta^{\tilde{E}}_{u_V}$  \\ 
     &$\alpha'^{H}_{u_V}$, $R_{2,u}^H$ &$\alpha'^{E}_{u_V}$, $R_{2,u}^E$ & $\alpha'^{\tilde{H}}_{u_V}$, $R_{2,u}^{\tilde{H}}$ &$\alpha'^{\tilde{E}}_{u_V}$, $R_{2,u}^{\tilde{E}}$\\ \hline
    \multirow{2}{*}{$\bar u$} & $N^{H}_{\bar u}$,$\alpha^{H}_{\bar u}$,$\beta^{H}_{\bar u}$,& \multirow{2}{*}{$R^{E}_{\rm{sea}}$} & $N^{\tilde{H}}_{\bar u}$,$\alpha^{\tilde{H}}_{\bar u}$,$\beta^{\tilde{H}}_{\bar u}$,& \multirow{2}{*}{$R^{\tilde{E}}_{\rm{sea}}$}   \\
    & $\alpha'^{H}_{\rm{sea}}$,$b_{\rm{sea}}^H$,$R_{2,u}^H$ & &$\alpha'^{\tilde{H}}_{\rm{sea}}$,$b_{\rm{sea}}^{\tilde{H}}$,$R_{2,u}^{\tilde{H}}$ &\\\hline
    \multirow{2}{*}{$d_V$}& $N^{H}_{d_V}$,$\alpha^{H}_{d_V}$,$\beta^{H}_{d_V}$, & $N^{E}_{d_V}$,$\alpha^{E}_{d_V}$,$\beta^{E}_{d_V}$, & $N^{\tilde{H}}_{d_V}$,$\alpha^{\tilde{H}}_{d_V}$,$\beta^{\tilde{H}}_{d_V}$, & $N^{\tilde{E}}_{d_V}$,$\alpha^{\tilde{E}}_{d_V}$,$\beta^{\tilde{E}}_{d_V}$ \\
    & $\alpha'^{H}_{d_V}$, $R_{2,d}^H$ & $\alpha'^{E}_{d_V}$, $R_{2,d}^E$ & $\alpha'^{\tilde{H}}_{d_V}$, $R_{2,d}^{\tilde{H}}$ & $\alpha'^{\tilde{E}}_{d_V}$, $R_{2,d}^{\tilde{E}}$\\ \hline
    \multirow{2}{*}{$\bar d$}& $N^{H}_{\bar d}$,$\alpha^{H}_{\bar d}$,$\beta^{H}_{\bar d}$, & \multirow{2}{*}{$R^{E}_{\rm{sea}}$} & $N^{\tilde{H}}_{\bar d}$,$\alpha^{\tilde{H}}_{\bar d}$,$\beta^{\tilde{H}}_{\bar d}$, & \multirow{2}{*}{$R^{\tilde{E}}_{\rm{sea}}$} \\
    &$\alpha'^{H}_{\rm{sea}}$,$b_{\rm{sea}}^H$,$R_{2,d}^H$ & &$\alpha'^{\tilde{H}}_{\rm{sea}}$,$b_{\rm{sea}}^{\tilde{H}}$,$R_{2,d}^{\tilde{H}}$ & \\\hline
    \multirow{2}{*}{$g$} &$\color{blue} N^{H}_{g}$,$\color{blue}\alpha^{H}_{g}$,$\beta^{H}_{g}$,$\color{blue}\alpha'^{H}_{g}$ & \multirow{2}{*}{$R^{E}_{\rm{sea}}$} & $ N^{\tilde{H}}_{g}$,$\alpha^{\tilde{H}}_{g}$,$\beta^{\tilde{H}}_{g}$,$\alpha'^{\tilde{H}}_{g}$ & \multirow{2}{*}{$R^{\tilde{E}}_{\rm{sea}}$} \\
    & $\color{blue}b_{g}^H$, $\color{blue}R_{g,2}^H$,$\color{blue}R_{g,4}^H$ & & $b_{g}^{\tilde{H}}$, $R_{g,2}^{\tilde{H}}$ &\\
      \hline \hline
    \end{tabular}
    \caption{Summary of GUMP GPD parameters. The 6 parameters in blue are free parameters to be fitted in this work, whereas the others are fixed from previous work~\cite{Guo:2023ahv}. The $E$ and $\tilde{E}$ GPDs of $\bar u,\bar d$ and $g$ are set to be proportional to the corresponding $H$ and $\tilde{H}$ GPDs with the ratio $R_{\rm{sea}}^E$ and $R_{\rm{sea}}^{\tilde{E}}$ due to the lack of constraints on them. More about their definitions can be found in~\cite{Guo:2023ahv}. We note that an extra parameter $N_{\rm{amp}}$ for the normalization of the amplitude is used in the fit, besides these GPD parameters. More details will be discussed in the next subsection.}
    \label{tab:parasum}
\end{table}

With these extra constraints, the ansatz for the four GPD species $\{ H, \tilde{H}, E, \tilde{E} \}$ of the five different parton species $\{ u, \bar{u}, d, \bar{d}, g \}$ were given in total by around sixty parameters, as summarized in Table. \ref{tab:parasum}, the majorities of which have been fitted in the previous work~\cite{Guo:2023ahv}.
In this work, we will fix those quark parameters as the background and focus on determining the gluon GPDs from experimental data of DV$J/\psi$P. Besides, we add extra gluon PDF constraints in the small-$x$ region from a recent JAM global analysis \cite{Cocuzza:2022jye} to fix the forward limit of the gluon GPDs to match the kinematics of the $J/\psi$ electroproduction measurement.


\subsection{Fitting to $J/\psi$ electroproduction with GUMP}
\label{subsec_jp_gump}

We now turn to the implementation of deeply virtual $J/\psi$ production in the GUMP framework. The parameters we need to determine for the gluon GPDs are the ones given in the eq. (\ref{mom_ans}), namely, $N_g,\alpha_g,\beta_g,\alpha'_g,$ and $b_g$. Besides, we have $R_{\xi^2}^g$ and $R_{\xi^4}^g$ for the $\xi$-dependence. However, since the $J/\psi$ electroproduction measurements are mostly in the small-$x_B$ region, which constrain the large-$x$ behavior of the gluon GPD very poorly, we fixed $\beta_g=7$ from the previous analysis, where $\beta_g$ accounts for the large-$x$ behavior of gluon PDF. Thus, we are left with 6 parameters to be determined from the small-$x$ gluon PDFs and the $J/\psi$ electroproduction measurement.

However, an extra subtlety appears in the meson DA of $J/\psi$. In the last section, we showed that a matching between the NRQCD and DVMP framework can be achieved when choosing the DA $\Phi_{\rm{NR}}(z)=\delta\left(z-\frac{1}{2}\right)$. This approximation shall receive further non-relativistic corrections, whereas the knowledge of the actual meson DA is rather limited, especially for heavy meson like $J/\psi$. There have been recent developments on the calculation of heavy meson DA with lattice simulation from first principle, which was applied to the $D$ meson with heavy-light flavor~\cite{Han:2024min}. However, the meson DA for heavy-heavy flavor remains a question. For simplicity, we will use the asymptotic DA $\Phi_{\rm{asym}}(z)=6z(1-z)$ in the analysis and this will cause a potential mismatch factor, which will be $2/3$ at leading order since $\int \text{d} z \ \Phi_{\rm{asym}}(z)/(1-z)=3$ instead of $2$ for the non-relativistic DA $\Phi_{\rm{NR}}(z)$. At NLO and beyond, this mismatch factor is not known, and thus we will parameterize it and determine it from the experiment. Consequently, the amplitude can be written as,
\begin{align}
\label{eq:dvmpamp}
\begin{split}
    \mathcal{A}_{\rm{Hyb.}} =\sum_{i=q,\bar q, g,\cdots} \frac{ e_q f_{J/\psi} }{N_C} \frac{Q}{Q^2+M_{J/\psi}^2}\int_0^{1}\text{d}z \int\limits_{-1}^1 \dd{x}& C^i (x, \xi,z,Q,\mu_R,\mu_{f,\rm{GPD}},\mu_{f,\rm{DA}}) \\& \times F^i (x, \xi, t,\mu_{f,\rm{GPD}}) N_{\rm{amp}} \Phi_{\rm{asym}}(z,\mu_{f,\rm{DA}}) \ ,
\end{split}
\end{align}
where $N_{\rm{amp}}$ stands for the mismatch factor after the substitution of asymptotic DA $\Phi_{\rm{asym}}$.

These $7$ parameters can then be fit to the small-$x$ gluon PDFs taken from \cite{Cocuzza:2022jye} and the DV$J/\psi$P cross-section~\cite{H1:2005dtp}. Before moving on to the actual fit, we note that the above amplitude can be extended to NLO by virtue of the previous development of the NLO Wilson coefficient and GPD evolution approaches in the literature \cite{Muller:2013jur, Duplancic:2016bge,Curci:1980uw,Belitsky:1998uk,Kumericki:2007sa,Cuic:2023mki} which we will not repeat here. Eventually, one can write the amplitude in the form of \footnote{Note that a factor of $C_F$ has been pulled out of the Wilson coefficient $C_k$ in the conformal moment space to match with the convention in the literature, see for instance~\cite{Cuic:2023mki}.}
\begin{align}
    \mathcal{A}_{\rm{Hyb.}} =& \frac{ e_q f_{J/\psi} C_F}{N_C} \frac{Q}{Q^2+M_{J/\psi}^2} \sum_{i = u, \, \bar{u}, \, ...} \, \frac{1}{2i}\int\limits_{c-i \infty}^{c + i \infty} \dd{j} \xi^{-j-1} \left[ i + \textrm{tan} \left( \frac{\pi j}{2} \right) \right]   \\
     &\times\Bigg[ C^{i, LO}_k E^{LO}_{kj}  F^i_j (\xi, t)+ C^{i, NLO}_k E^{LO}_{kj}  F^i_j (\xi, t) + C^{i, LO}_k  E^{NLO}_{jk} F^i_j (\xi, t) \Bigg]  , \notag
\end{align}
where we have only kept terms up to NLO accuracy. The scale-dependence is suppressed in the above expression for clarity. More details regarding the scale-dependence and the implementation of the GPD evolution can be found in the appendix.

We also note that the skewness variable $\xi$ will be modified in the presence of vector meson mass $M_{J/\psi}$, which reads
\begin{align}
    \xi = \frac{\tilde{x}_B}{2-\tilde{x}_B} ,
\end{align}
in terms of the meson mass corrected Bjorken variable
\begin{align}
    \tilde{x}_B = \frac{M_{J/\psi}^2 + Q^2}{W^2 + Q^2},
\end{align}
with $Q^2$ the photon virtuality and $W^2$ the center-of-mass energy squared for the scattering of the virtual photon on the hadron. NLO corrections in the valence quark sector will not be considered in this work because in $J/\psi$ production the valence contribution would require intrinsic charm content in the proton, which we do not consider for now.

At the end of the subsection, we note again that a complete matching of the DVMP and the NRQCD frameworks at NLO will be crucial for the proper understanding of the systematical uncertainties in the analysis and will be left to a separate work.


\subsection{Input PDFs and differential cross-sections and fit results}
\label{subsec_res}

Having established the inputs and framework, we now turn to the results. We sample 9 points of gluon PDF from the globally extracted set \cite{Cocuzza:2022jye} at $\mu = 2 \ \textrm{GeV}$ in the region $x \in \left[10^{-4},10^{-3}\right]$, and take the 17 points of differential cross-sections from H1~\cite{H1:2005dtp} with $\left< Q^2 \right> \in \left[7.0, 22.4\right] \ \textrm{GeV}^2, -t \in \left[ 0.04, 0.64\right] \ \textrm{GeV}^2$, and $\tilde{x}_B \in \left[9\times 10^{-4},6\times 10^{-3}\right]$ to fit the 7-parameter set $\left\{ N^g, \alpha^g, \beta^g, R_{\xi^2}^g, R_{\xi^4}^g, N_{\textrm{amp}} \right\}$. The best-fit parameters are obtained with the \texttt{iminuit} interface of \texttt{Minuit2}~\cite{iminuit,James:1975dr} as the minimizer, yielding $\chi^2/$d.o.f.$\approx 1.53$, which are listed with their statistical uncertainties from \texttt{Minuit2} in Table~\ref{tab_params}. 

\begin{table}[h] 
    \def\arraystretch{1.25}
    \centering
    \begin{tabular}{>{\centering\arraybackslash} m{2cm}|>{\centering\arraybackslash} m{4cm} |>{\centering\arraybackslash} m{4cm}}
    \hline
    \hline
      Parameter & Best-Fit Value & Statistical Uncertainty \\ \hline\hline
      $N^g$ & 1.78 & 0.21 \\ \hline
      $\alpha^g$ & $1.100$ & 0.016 \\ \hline
      $\alpha'^g$ & 0.02 & 0.04 \\ \hline
      $R_{\xi^2}^g$ & $-0.63$ & 0.30 \\ \hline
      $R_{\xi^4}^g$ & $0.26$ & 0.10 \\ \hline
      $b^g$ & 1.84 & 0.12 \\ \hline
      $N_{\rm{amp}}$ & 1.42 & 0.14 \\ \hline\hline
    \end{tabular} 
    \caption{The best-fit values for the seven parameters are listed along with their statistical uncertainties from \texttt{Minuit2} with $\chi^2/$d.o.f.$\approx1.53$.}
    \label{tab_params}
\end{table}

\begin{figure}[ht]
\centering
\includegraphics[width= 0.5 \linewidth]{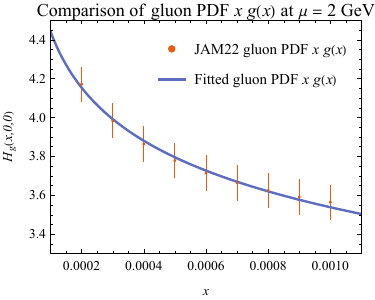}
\caption{Comparison of the fitted forward limit of the GPDs to the PDF data points taken from the JAM22 global analysis~\cite{Cocuzza:2022jye} in the region of small $x \sim 10^{-4} - 10^{-3}$ at the reference scale $\mu_0 = 2$ GeV. }
\label{FIG:smallxpdf}
\end{figure}

In Fig.~\ref{FIG:smallxpdf}, we compare the fitted PDFs to the samples of PDFs taken from the JAM22 global analysis~\cite{Cocuzza:2022jye}, where the two forward parameters $N_g$ and $\alpha_g$ appear to describe the small-$x$ gluon PDFs well. Extending to the region with larger $x$ would require a more flexible parameterization of the gluon PDFs, which will be left to future work with more inputs. In this work, we focus on the small-$x$ gluon PDF and its impact on the DV$J/\psi$P cross-section data. Thus, the two parameters are adequate for this purpose.

\begin{figure}[ht]
\centering
\includegraphics[width= 0.8 \linewidth]{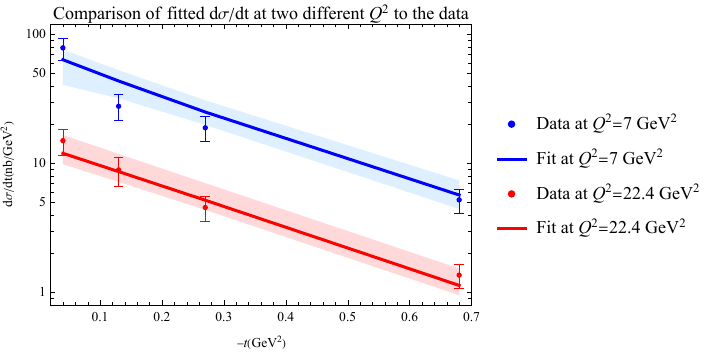}
\caption{Comparison of the fitted differential cross-section calculated with eq.(\ref{r_rat_xsec}) and the best-fit parameters in table \ref{tab_params}, as a function of $t$ against the H1 data~\cite{H1:2005dtp} for various $Q^2$ and $x_B$ ranging from $1.3\times10^{-3}$ up to $3.2\times 10^{-3}$. The bands are calculated by variation of scale with $\mu^2 \in \left[1/2,2\right] \times \mu_F^2$.}
\label{FIG:tdep}
\end{figure}

\begin{figure}[ht]
\centering
\includegraphics[width= 0.8 \linewidth]{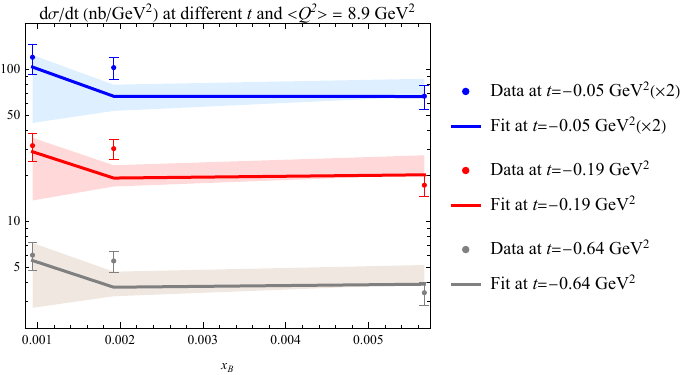}
\caption{Similar to the previous plot, we present the comparison of the fitted differential cross-section calculated with eq.(\ref{r_rat_xsec}) and the best-fit parameters in table \ref{tab_params} as a function of $x_B$ against the H1 data~\cite{H1:2005dtp} for various values of $t$, with $\left<Q^2\right> = 8.9 \rm{~GeV}^2$. Again, the bands are calculated by variation of scale with $\mu^2 \in  \left[1/2,2\right] \times \mu_F^2$. The multipliers in the parentheses have been multiplied to the cross-sections to avoid overlaps. }
\label{FIG:xdep}
\end{figure}

In Fig.~\ref{FIG:tdep} and Fig.~\ref{FIG:xdep}, we show the differential cross-section for DV$J/\psi$P as a function of $t$ and $x_B$ respectively from the H1 data~\cite{H1:2005dtp}, comparing to the best-fit values. We observe large bands obtained by scale variation of $\mu^2$ with a factor of two, indicating potentially large corrections that are higher order of $\alpha_s$, i.e., NNLO and beyond. We note that there are measurements of differential cross-section of DV$J/\psi$P by ZEUS~\cite{ZEUS:2004yeh} as well. However, some slight tensions are noticed when fitting the two measurements altogether, and thus we pick the H1 data which has more coverage in the high-$Q^2$ region. Additionally, there are a few more points for discussion. First, we found the NLO corrections are indeed required to describe the data --- the LO calculation will not reproduce the correct $x_B$- and $Q$-scaling of the data, no matter of the input GPD. The same applies to the leading mass corrections of $M_{J/\psi}$ as well, where the data are found to scale according to $1/\left(Q^2+M_{J/\psi}^2\right)^n$ which simply cannot be described by the $1/Q^{n}$-scaling with the DVMP framework. 

\begin{figure}[ht]
\centering
\includegraphics[width= 0.8\textwidth]{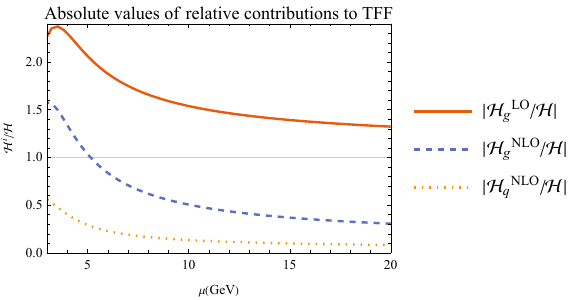}
\caption{The relative contributions of LO gluon $\mathcal{H}_g^{\rm{LO}}$, NLO gluon $\mathcal{H}_g^{\rm{NLO}}$ and NLO quark $\mathcal{H}_q^{\rm{NLO}}$ contributions to the full NLO TFF $\mathcal{H}$ decomposed according to eq. (\ref{eq:Hdecomp}) at $x_B=0.005$ and $t=-0.05$ GeV. Note that at LO there is no quark contribution. The ratios are complex numbers in general, and their absolute values are presented here.}
\label{FIG:tff}
\end{figure}

In fact, strong cancelation between the LO gluon and NLO gluon contributions is found in the TFFs and final cross-sections. In Fig. \ref{FIG:tff}, we present the relative contributions of LO gluon $\mathcal{H}_{g}^{\rm{LO}}$, NLO gluon $\mathcal{H}_{g}^{\rm{NLO}}$ and NLO quark $\mathcal{H}_{q}^{\rm{NLO}}$ to the TFF $\mathcal{H}$ respectively at one reference kinematic point and their evolution with the reference scale $\mu$. Note that the NLO here excludes the LO contributions so that we write:
\begin{equation}\label{eq:Hdecomp}
\mathcal{H}=\mathcal{H}_{g}^{\rm{LO}}+\mathcal{H}_{g}^{\rm{NLO}}+\mathcal{H}_{q}^{\rm{NLO}}\ ,
\end{equation}
It appears that the quark fields always have relatively small contributions, whereas the gluonic contributions from NLO are rather comparable to the LO ones, as shown in the figure on the right. Although the NLO corrections seem to be suppressed for increasing $\mu$, as expected for a perturbative expansion.

These observations, together with the strong scale dependence shown in Fig.~\ref{FIG:tdep} and Fig.~\ref{FIG:xdep}, imply that the perturbative expansion in $\alpha_s$ is not as well-controlled in the case of meson production as in, for example, DVCS, where the LO calculation is sufficient to produce a reasonably good fit even when including both low-$Q^2$ JLab data and small-$x_B$ HERA data. Such observations are consistent with what has been found in the literature~\cite{Chen:2019uit,Flett:2021ghh,Eskola:2022vpi} in various frameworks, even including the light-meson production~\cite{Muller:2013jur,Cuic:2023mki}. Therefore, further studies on the perturbative corrections as well as non-relativistic corrections for heavy meson productions are crucial for reducing the systematic uncertainties. We also notice that recently it has been found that the scale dependence in the exclusive photo-production of $J/\psi$ can be reduced by resumming higher-order QCD corrections~\cite{Flett:2024htj}. In the case of lepto-production here, we find that the best-fit predictions for DV$J/\psi$P are closer to the data as $Q^2$ increases, which could be an indication of better convergence at large $Q^2$. However, there is not many existing high-precision data at large $Q^2$, which should be one vital task of the future EIC~\cite{AbdulKhalek:2021gbh}.

As a cross-check, we comment that the results obtained in this work are rather consistent with \cite{Chen:2019uit} in the NRQCD framework at NLO, even though only simple GPD models are used there. Comparing to the results therein, we found similar large scale-dependence as well as strong cancelation between LO and NLO contributions. Though the theoretical values agree better with the data in this work, by virtue of the extra flexibility of $\xi$-dependence in the GPD parameterization that has been fitted to the data, as well as the full NLO evolution of GPD that potentially helps. This agreement indicates that the hybrid framework in this work captures the main feature of NLO corrections, whereas more careful studies of their difference will be left to future work.

\subsection{Extracted gluon GPDs and other relevant observables}

\begin{figure}[ht]
\centering
\includegraphics[width= 0.95\linewidth]{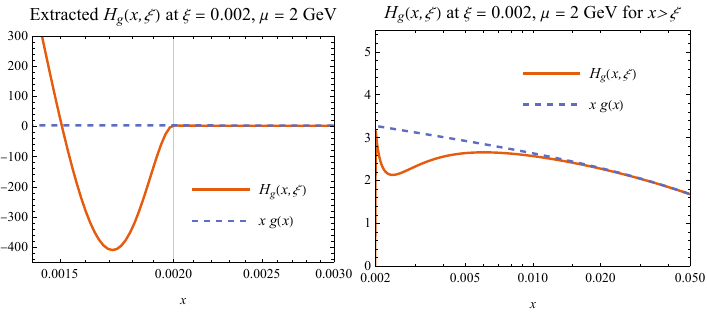}
\caption{Extracted gluon GPD $H_g(x,\xi,t=0)$ at $\xi=0.002$ and reference scale $\mu=2$ GeV. The left plot shows the oscillating behavior in the DA-like region, and the right plot shows that GPDs reduce to PDFs when $x\gg \xi$ in the PDF-like region with $t=0$.}
\label{FIG:gGPD}
\end{figure}

Finally, we also discuss the extracted gluon GPDs themselves. It is well-known that processes like DVCS and DVMP do not provide enough constraints on the shape of the GPDs themselves, as discussed in the context of shadow GPD~\cite{Bertone:2021yyz}. Therefore, all extraction of the GPDs from those process alone without extra off-forward input cannot fully determine the shape of GPDs. Despite that, we present the shape of gluon GPDs obtained from this extraction for reference in Fig. \ref{FIG:gGPD}. In the right figure, it is shown that the gluon GPDs approaches the gluon PDFs in the limit $x\gg\xi$ as expected. On the other hand, strong oscillation is shown in the DA-like region of GPD when $-\xi<x<\xi$, as shown in the left figure. This behavior has been extensively discussed in the previous work for the quark GPDs~\cite{Guo:2023ahv}, which can be considered the artifacts of the conformal moment parameterization. More careful treatment can be done to tune the GPDs in the DA-like region. However, in the small-$\xi$ limit, the DA-like region becomes singular itself and therefore, those behaviors might not have too much physical impact anyway.

\begin{figure}[ht]
\centering
\includegraphics[width= 0.7 \linewidth]{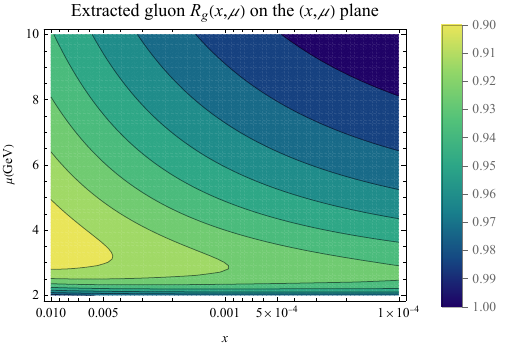}
\caption{Skewness ratio $R_g(x,\mu)$ from the extracted gluon GPD shown on the $(x,\mu)$ plane in the region covered by the data and extrapolated to large $\mu$.}
\label{FIG:Rdep}
\end{figure}

Another interesting feature worth discussing is the $\xi$-dependence of gluon GPDs obtained in this framework. Due to the lack of input of the $\xi$-dependence, it is commonly assumed that the amplitudes can be approximated by the corresponding PDFs $H_g(x,\xi=0)$ up to some factors for skewness corrections. However, the skewness-dependence is likely to be more complicated and more importantly, it will evolve with the renormalization scale $\mu$. Thus, our extraction here provides an opportunity to study such effects, even though we note that it will be model-dependent. In Fig. \ref{FIG:Rdep}, we present the $x$- and $\mu$-dependence of the skewness ratio or the so-called $R$-ratio defined by:
\begin{equation}
    R_{g}(x,\mu)\equiv \frac{H_{g}(x,\xi=x,0,\mu)}{H_{g}(x,0,0,\mu)} \ .
\end{equation}
As clearly shown in Fig. \ref{FIG:Rdep}, as $x\to0$ and $\mu\to \infty$, the skewness ratio somewhat stabilizes around the value $R_g(x,\mu)\sim 1$ that approaches unity. However, for majority values of $x$ and $\mu$, the $R_g(x,\mu)$ deviates from unity. This indicates the skewness effect even for the kinematics of the H1 data, which has $\xi \sim 10^{-4}$. We note that in another recent work~\cite{Cuic:2023mki}, the DVCS and DVMP measurements at HERA including only light meson productions have been studied at NLO in the GPD framework, based on a similar conformal moment framework~\cite{Muller:2013jur}. The skewness ratio is found to be around $1.1$ therein, which is larger than one. Nevertheless, we note that the extraction of the ratio $R_g(x,\mu)$ is rather model-dependent. Similar skewness effects have also been seen in small-$x$ frameworks \cite{Toll:2012mb,Mantysaari:2016jaz} and it appears that the off-forward physics separating GPDs from their PDF limits is non-trivial even for very near-forward scattering.

In the end, we comment on the fitting parameter and the model/extraction-dependence of the results. In this work, we aim to bring together the forward small-$x$ gluon PDFs and the DV$J/\psi$P cross-section data in the GPD framework at the NLO. In the forward region, the two parameters $N_g$ and $\alpha_g$ are flexible enough for the small-$x$ gluon PDF. More flexibility is definitely needed to extend the analysis to a larger region to include the larger-$x$ sector of gluon PDF, as well as lattice calculations of the gluon gravitational form factors~\cite{Pefkou:2021fni,Hackett:2023rif}. However, at this point, we still have to carefully balance the sophistication of our ansatz in the forward limit against the availability of off-forward data inputs.

As for the off-forward parameters, two parameters for the $t$-dependence, $b_g$ and $\alpha'_g$ and another two parameters for the $\xi$-dependence, $R_{\xi^2}^g$ and $R_{\xi^4}^g$ have been used in the ansatz. Although the actual $t$-dependence and $\xi$-dependence can be much more complicated than this, we do see the four parameters reproduce the $t$- and $x_B$-dependence of the data well. One thing worth noting is the fixed quark parameters throughout this analysis, as we assume that the quark contributions are suppressed, at least in the limit $\mu \to \infty$. While our results in Fig.~\ref{FIG:tff} seem to confirm such behaviors, it should be worthy to consider a simultaneous fit of quark and gluon distributions from combined input to further justify this assumption, which will be left to a future work. The last parameter is the normalization of the amplitude $N_{\rm{amp}}$ which depends on the choice of the meson DA $\Phi(z)$. From the fit, we obtain $1.42\pm0.14$ which is reasonably close to the LO values $2/3$, noting that it strongly depends on the choice of the factorization scale. Since the results receive potentially large perturbative corrections as discussed before, this factor should be more carefully studied by a complete matching of the two frameworks at NLO. 

\section{Conclusions and outlook}
\label{sec_con}

To conclude, we study the gluon GPDs with DV$J/\psi$P process, employing a hybrid framework that combines NLO collinear factorization with LO heavy $J/\psi$ mass corrections from NRQCD. We find that the differential cross-section of DV$J/\psi$P measurement by H1~\cite{H1:2005dtp} can be described with the gluon GPDs after including the NLO perturbative corrections and LO heavy mass corrections, when simultaneously fitting to the forward gluon PDFs obtained from global analysis~\cite{Cocuzza:2022jye}. We study the scale-dependence as well as the contributions of different order to the amplitudes, and find significant contributions from the gluon at NLO that cancel the LO contributions, as well as strong scale dependence even at NLO, indicating large perturbative corrections that should be further studied. These observations agree with the previous work that studies the exclusive $J/\psi$ lepto-production in the NRQCD framework at NLO with simple GPD models~\cite{Chen:2019uit}.

Additionally, we present the gluon GPDs as well as the skewness ratio based on the extraction of this work, though the extraction of GPDs based on such exclusive meson productions alone suffers from the inverse problem and therefore would depend on the specific model. Our extracted skewness ratio appears to be consistent with another extraction from combined DVCS and DVMP analysis with light meson productions at NLO~\cite{Cuic:2023mki}, whereas similar skewness effects have been seen in other frameworks as well~\cite{Mantysaari:2016jaz}.

For future development, more careful studies to match the collinear DVMP and NRQCD framework are crucial to systematically examine the NLO corrections as well as to better understand the effect of the mismatch factor $N_{\rm {amp}}$. Besides, including other light-meson production as well as DVCS measurements altogether at NLO would allow us to better determine the quark and gluon GPDs simultaneously, so that their contributions, especially at NLO, can be better understood. In addition, the GUMP framework also enables us to extend to the larger-$x$ region, so that a global analysis that includes gluon PDFs in larger-$x$ region and gluonic form factors from lattice~\cite{Pefkou:2021fni,Hackett:2023rif} can be achieved. 

\section*{Acknowledgments}

We thank Feng Yuan for many useful discussions and comments related to the subject of this paper. We also thank K. Kumeri\v{c}ki for the correspondence on GPD evolution, as well as his private notes on this. This research also benefits from the open-source Gepard package~\cite{gepard}. This research is supported by the U.S. Department of Energy, Office of Science, Office of Nuclear Physics, under contract number DE-SC0020682 and under award number DE-AC05-06OR23177, and the Center for Nuclear Femtography, Southeastern Universities Research Association, Washington D.C. The authors also acknowledge partial support by the U.S. Department of Energy, Office of Science, Office of Nuclear Physics under the umbrella of the Quark-Gluon Tomography (QGT) Topical Collaboration with Award DE-SC0023646.
\appendix

\section{Setup for the NLO hard coefficients and evolution}

In this appendix, we present the details of the implementation of the NLO hard coefficients and evolutions. In eq. (\ref{eq:conformalsum}), we show that the GPDs are expanded in terms of their conformal moments. As shown in eq. (\ref{eq:dvmpamp}), the DVMP amplitudes are given by a double convolution of GPDs with the Wilson coefficients $C^g (x, \xi,z,Q,\mu)$ as well as the meson distribution amplitude $\Phi_{\rm{asym}}(z,\mu)$. In moment space, the double convolution will be turned into a double summation in the form of:
\begin{align}
\begin{split}
    \mathcal{A}_{\rm{Hyb.}} =\frac{ e_q f_{J/\psi} C_F}{N_C} \frac{Q}{Q^2+M_{J/\psi}^2}\sum_{i=q,\bar q, g}  \sum_{j=0}^\infty\sum_{l=0}^\infty[1\mp(-1)^j] \xi^{-j-1} &C^i_{jl}(Q,\mu_R,\mu_{f,\rm{GPD}},\mu_{f,\rm{DA}}) \\&\times F_{j}^i(\xi,t,\mu_{f,\rm{GPD}}) \Phi_{\it{l}}(\mu_{f,\rm{DA}}) 
 \ ,
 \end{split}
\end{align}
where $F_{j}^i(\xi,t,\mu_{f,\rm{GPD}})$ and $\Phi_{\it{l}}(\mu_{f,\rm{DA}})$ are the GPD and DA in conformal moment space. The sign $\mp$ is $-$ for vector GPDs and $+$ for axial-vector GPDs. The asymptotic DA $\Phi^{\rm{asym}}(z,\mu)$ contains the leading conformal moment only: $\Phi^{\rm{asym}}_l(\mu)=\delta_{l0}\Phi^{\rm{asym}}_0(\mu)$. 

For simplicity, we set all the factorization and renormalization scales to be the same: $\mu=\mu_{f,\rm{DA}}=\mu_{f,\rm{GPD}}=\mu_R$, though the choice of $\mu$ appears to be a bit subtle. It has been suggested in \cite{Chen:2019uit} that the scale of a single quark rather than that of the quark pair can be chosen to suppress the higher-order perturbative corrections in the high-energy limit. Namely, the leading double logs are observed to be in the form of $\log\left(\frac{Q^2+M_{J/\psi}^2}{4\mu_F^2}\right)\log\left(\xi\right)$. In the DVMP framework, the large logs are in the form of $\log(Q^2/\mu^2)$, and thus we set $\mu=Q$ accordingly to suppress the higher-order perturbative corrections. We note that the results will strongly depend on the choice of scale, which shall be further studied.
\begin{align}
    \mathcal{A}_{\rm{Hyb.}} =\frac{ e_q f_{J/\psi}C_F }{N_C} \frac{Q}{Q^2+M_{J/\psi}^2}\sum_{i=q,\bar q, g}  \sum_{j=0}^\infty [1\mp(-1)^j]\xi^{-j-1} C^i_{j0} F_{j}^i(\xi,t,\mu) \Phi^{\rm{asym}}_0(\mu) 
 \ ,
\end{align}
where the GPD and DA are evolved to the scale $\mu$ accordingly. Note that the asymptotic DA corresponds to the asymptotic behavior of DA in the $\mu_f \to \infty$ limit, and thus its scale dependence is trivial such that we can write it as $\Phi^{\rm{asym}}_0$ for simplicity, which could be recovered if necessary~\footnote{The asymptotic DA is invariant under LO evolution, whereas the NLO evolution effects are negligible. Here we ignore the NLO evolution of DA which is also partially accounted in the $N_{\rm{amp}}$, so the scale-dependence of DA becomes trivial.}. On the other hand, the evolution of GPD is more complicated, which can be written in terms of the evolution operator $E^{ii'}_{jk}(\xi,\mu,\mu_0)$ as,
\begin{align}
    F^{i}_{j}(\xi,t,\mu)=\sum_{i'=q,\bar q,g}\sum_{k=0}^j\left[\frac{1\mp(-1)^k}{2}\right] E^{ii'}_{jk}(\xi,\mu,\mu_0) F^{i'}_{k}(\xi,t,\mu_0)\ .
\end{align}
Note that the evolution operator includes non-trivial mixing in both the flavor ($i=q,\bar q,g$) space and the conformal moment ($j$) space. Therefore, we can write the amplitude as,
\begin{align}
\label{eq:ampdoublesum}
\begin{split}
       \mathcal{A}_{\rm{Hyb.}} =\frac{ e_q f_{J/\psi}C_F }{N_C} \frac{Q}{Q^2+M_{J/\psi}^2}&\sum_{i,i'=q,\bar q, g}  \sum_{j=0}^\infty\sum_{k=0}^j [1\mp(-1)^j]\xi^{-j-1} C^i_{j0} \\&\times \left[\frac{1\mp(-1)^k}{2}\right] E^{ii'}_{jk}(\xi,\mu,\mu_0)
    F_{k}^{i'}(\xi,t,\mu_0)\Phi^{\rm{asym}}_0
 \ .
\end{split}
\end{align}
Again, the signs $\mp$ are $-$ for vector GPDs and $+$ for axial-vector GPDs. The Wilson coefficients $C_{j0}^i$ and the evolution operators $E_{jk}^i(\xi,\mu,\mu_0)$ and $E_{\Phi_0}(\mu,\mu_0)$ are known to the NLO~\cite{Curci:1980uw,Belitsky:1998uk,Kumericki:2007sa}, whereas the GPD $F_{k}^i(\xi,t,\mu_0)$ and the DA $\Phi^{\rm{asym}}_0(\mu_0)$ are parameterized at the reference scale $\mu_0$. Thus, the amplitude can in principle be evaluated by performing the double summation, which, however, is more involving than it appears.

\subsection{Resummation of moments with Mellin-Barnes integral}

As discussed in the main text as well as many previous developments on conformal moment construction of GPDs, the resummation in moment space is typically divergent that requires proper analytic treatment. The common one in the literature is the Mellin-Barnes integral, which basically states:
\begin{equation}
    \sum_{j=j_0}^\infty (-1)^j f_j  =\frac{1}{2 i} \int_{c-i\infty}^{c+i\infty} \frac{1}{\sin(\pi j)} f_j \ ,
\end{equation}
where $c$ is a real number that satisfies $j_0-1<c<j_0$. Requiring that $f_j$ is analytic when $\text{Re}j> c$ and falls fast enough at infinite, the identity can be proven simply with residue theorem after analytical continuation. This also indicates similar variants:
\begin{align}
    \sum_{j=j_0}^\infty [1-(-1)^j] f_j  &= -\frac{1}{2 i}\int_{c-i\infty}^{c+i\infty}\tan(\pi j/2) f_j \ , \\
    \sum_{j=j_0}^\infty [1+(-1)^j] f_j  &= \frac{1}{2 i}\int_{c-i\infty}^{c+i\infty} \cot(\pi j/2)f_j \ ,
\end{align}
which can be used to resum the double summation in eq. (\ref{eq:ampdoublesum}). Also note that since the second summation in $k$ only sum over $0\le k \le j$, not to infinity, one can reshuffle the double summation to show that
\begin{equation}
 \sum_{j=0}^\infty\sum_{k=0}^j F_{jk} = \sum_{k=0}^\infty\sum_{j=k}^\infty F_{jk} = \sum_{j=0}^\infty\sum_{k=j}^\infty F_{kj} \ .
\end{equation}
Therefore, we have alternatively for eq. (\ref{eq:ampdoublesum})~\cite{Kumericki:2007sa}:
\begin{align}
\label{eq:ampdoublesum2}
\begin{split}
       \mathcal{A}_{\rm{Hyb.}} =\frac{ e_q f_{J/\psi} C_F}{N_C} \frac{Q}{Q^2+M_{J/\psi}^2}&\sum_{i,i'=q,\bar q, g}  \sum_{j=0}^\infty\sum_{k=j}^\infty [1\mp(-1)^k]\xi^{-k-1} C^i_{k0} \\&\times \left[\frac{1\mp(-1)^j}{2}\right] E^{ii'}_{kj}(\xi,\mu,\mu_0)
    F_{j}^{i'}(\xi,t,\mu_0) \Phi^{\rm{asym}}_0
 \ .
\end{split}
\end{align}
In both eqs. (\ref{eq:ampdoublesum}) and (\ref{eq:ampdoublesum2}), the second summation of $k$ should be performed before the $j$ summation. However, the meaning of $k$ has been switched due to the change of variable. In the former one in eq. (\ref{eq:ampdoublesum}), $k$ refers to the $k$th moment of the unevolved GPD, whereas in the latter one in eq. (\ref{eq:ampdoublesum2}), $k$ refers to the $k$th moment of the Wilson coefficients. Accordingly, the two different orders of performing the double summation will be referred to as the moment-evolution method and the coefficient-evolution method.

Although the two methods should produce the same results, both setups are needed in the actual implementation of NLO evolution. During the fit, one needs to change the parameters of GPDs and thus the GPD moments frequently. Thus, it will be much more efficient to pre-compute the evolved Wilson coefficients and sum them with the moments afterward. On the other hand, for the calculation of, e.g., best-fit GPDs, after the fit, the moments of GPDs will be fixed whereas different $x$ and $\xi$ will be put in. Then the moment-evolution method will be the more reasonable choice. Actually, we found it extremely difficult to perform the analysis otherwise --- both fitting with the moment-evolution method and computing GPDs with the coefficient-evolution method seem too slow for any practical purpose. More details will be discussed in the following subsection.

\subsection{Implementation and comparison of the two methods}

Finally, we discuss how the two methods are numerically implemented. We first note that the NLO evolution has been implemented in the \texttt{Gepard} package~\cite{gepard}. Our work is based on the existing \texttt{Gepard} development, and aims to provide a complete implementation of the NLO evolution including both methods. Before getting to the details, we first note that there are two useful properties of the evolution kernel for the implementation. First, the $\xi$-dependence of the evolution kernel can be factorized at least to the NLO as:
\begin{equation}
    E^{ii'}_{jk}(\xi,\mu,\mu_0) = \xi^{j-k} E^{ii'}_{jk}(1,\mu,\mu_0) \ .
\end{equation}
Therefore, we define $\bar E^{ii'}_{jk}(\mu,\mu_0)= E^{ii'}_{jk}(1,\mu,\mu_0)$ that is independent of $\xi$. Second, the evolution kernel can be split into diagonal and off-diagonal parts:
\begin{equation}
    \bar{E}^{ii'}_{jk}(\mu,\mu_0) = \delta_{kj}A^{ii'}_{j}(\mu,\mu_0) + B^{ii'}_{jk}(\mu,\mu_0)\ ,
\end{equation}
where the diagonal term $A^{ii'}_{j}(\mu,\mu_0)$ shows up at LO, while the $B^{ii'}_{jk}(\mu,\mu_0)$ is NLO and beyond. The double summation involving the diagonal term reduces to a simple single summation, which can be treated in the same way as in the LO case. Therefore, we focus on the off-diagonal piece associated with the $B^{ii'}_{jk}(\mu,\mu_0)$. 

In eq. (\ref{eq:ampdoublesum}) for the moment-evolution method, we have the double summation in the form of, \begin{align}
\begin{split}
  &\sum_{j=0}^\infty\sum_{k=0}^j [1-(-1)^j]\xi^{-j-1} C^i_{j0}\left[\frac{1-(-1)^k}{2}\right] E^{ii'}_{jk}(\xi,\mu,\mu_0)
    F_{k}^{i'}(\xi,t,\mu_0)\\
=&\sum_{j=0}^\infty\sum_{k=0}^j [1-(-1)^j]\xi^{-k-1} C^i_{j0}\left[\frac{1-(-1)^k}{2}\right] \left[\delta_{jk}A^{ii'}_{j}(\mu,\mu_0)+B^{ii'}_{jk}(\mu,\mu_0)\right]
    F_{k}^{i'}(\xi,t,\mu_0)\ ,
\end{split}
\end{align}
where we ignore the prefactors that are independent of $j$ and $k$ for clarity. Then the double summation of the off-diagonal term can be converted into the following double integral \cite{Zhang:2024djl}:
\begin{align}
\begin{split}
  &\sum_{j=1}^\infty\sum_{k=1}^j [1-(-1)^j]\xi^{-k-1} C^i_{j0}\left[\frac{1-(-1)^k}{2}\right] B^{ii'}_{jk}(\mu,\mu_0)
    F_{k}^{i'}(\xi,t,\mu_0) \\
    \to&~ \frac{1}{2i}\int_{c_j-i \infty}^{c_j+i \infty} \mathrm{d} j\left[i+\tan\left(\frac{\pi j}{2}\right)\right] \xi^{-j-1} C_{j}^i\int_{c_k-i \infty}^{c_k+i \infty} \frac{\mathrm{d} k}{4i}\cot\left(\frac{\pi k}{2}\right)\\
    &\qquad\qquad\qquad\qquad\qquad\qquad\left(\xi^{-k} B_{j, k+j}^{i i^{\prime}} \mathcal{F}_{k+j+2}^{i^{\prime}}-\xi^{j-k-1} B_{j, k+1}^{i i^{\prime}} \mathcal{F}_{k+1}^{i^{\prime}}\right),
\end{split}
\end{align}
where $c_j<2$ and $c_k<0$. In the above expression, we explicitly exclude the contributions from the $k=0$ terms which can be added to the results later. The benefit of doing so is to avoid the pole in the GPD moment $F_k$ as well as the poles in the anomalous dimensions in the evolution kernel. Those poles must be avoided so that the contour integral is properly defined. Similarly, for the coefficient-evolution method, the following double integral representation can be derived:
\begin{align}
\begin{split}
  &\sum_{j=1}^\infty\sum_{k=j}^\infty [1-(-1)^k]\xi^{-j-1} C^i_{k0}\left[\frac{1-(-1)^j}{2}\right] B^{ii'}_{kj}(\mu,\mu_0)
    F_{j}^{i'}(\xi,t,\mu_0) \\
    &\to \frac{1}{2i}\int_{c_j-i \infty}^{c_j+i \infty} \mathrm{d} j\left[i+\tan\left(\frac{\pi j}{2}\right)\right]\xi^{-j-1}\mathcal{F}_{j}^{i^{\prime}}\int_{c_k-i \infty}^{c_k+i \infty} \frac{\mathrm{d} k}{4i}\tan\left(\frac{\pi k}{2}\right)  C_{k+j+1}^i B_{k+j+1, j}^{i i^{\prime}} \ ,
\end{split}
\end{align}
with $c_j<1$ and $c_k<0$. Again, the contributions from $j=0$ terms are removed in the above expression and will be added back later to avoid the poles for the same reason.

Such techniques also apply to the case of axial-vector GPDs, which replaces $[1-(-1)^{j,k}]$ with $[1+(-1)^{j,k}]$, and we have:
\begin{align}
\begin{split}
  &\sum_{j=1}^\infty\sum_{k=1}^j [1+(-1)^j]\xi^{-k-1} C^i_{j0}\left[\frac{1+(-1)^k}{2}\right] B^{ii'}_{jk}(\mu,\mu_0)
    F_{k}^{i'}(\xi,t,\mu_0) \\
    \to&~ \frac{1}{2i}\int_{c_j-i \infty}^{c_j+i \infty} \mathrm{d} j\left[i-\cot\left(\frac{\pi j}{2}\right)\right]  \xi^{-j-1} C_{j}^i\int_{c_k-i \infty}^{c_k+i \infty}  \frac{\mathrm{d} k}{4i}\left\{\tan\left(\frac{\pi k}{2}\right)\xi^{j-k-1} B_{j, k+1}^{i i^{\prime}} \mathcal{F}_{k+1}^{i^{\prime}}\right.\\
    &\qquad\qquad\qquad\qquad\qquad\qquad\qquad\qquad\left.+\cot\left(\frac{\pi k}{2}\right)\xi^{-k} B_{j, k+j}^{i i^{\prime}} \mathcal{F}_{k+j}^{i^{\prime}}\right\}\ ,
\end{split}
\end{align}
with $c_j<2$ and $c_k<0$. And for the coefficient-evolution method, we have
\begin{align}
\begin{split}
  &\sum_{j=1}^\infty\sum_{k=j}^\infty [1+(-1)^k]\xi^{-j-1} C^i_{k0}\left[\frac{1+(-1)^j}{2}\right] B^{ii'}_{kj}(\mu,\mu_0)
    F_{j}^{i'}(\xi,t,\mu_0) \\
    &\to \frac{1}{2i}\int_{c_j-i \infty}^{c_j+i \infty} \mathrm{d} j\left[i-\cot\left(\frac{\pi j}{2}\right)\right]\xi^{-j-1}\mathcal{F}_{j}^{i^{\prime}}\int_{c_k-i \infty}^{c_k+i \infty} \frac{\mathrm{d} k}{4i}\tan\left(\frac{\pi k}{2}\right)  C_{k+j+1}^i B_{k+j+1, j}^{i i^{\prime}} \ ,
\end{split}
\end{align}
with $c_j<1$ and $c_k<0$. Noting that all $\tan\left(\pi j/2\right)$ are replaced by $-\cot\left(\pi j/2\right)$ just like the diagonal case, whereas the replacement of $\tan\left(\pi k/2\right)$ and $\cot\left(\pi k/2\right)$ is highly non-trivial due to the redefinition of variable in the second integral. These two methods have been numerically checked to produce the same amplitudes/GPDs with NLO evolution. The code has been updated online~\cite{Guo:2022gumpgit}.

\bibliographystyle{jhep}

\bibliography{refs.bib}

\providecommand{\href}[2]{#2}\begingroup\raggedright\begin{thebibliography}{100}

\bibitem{Muller:1994ses}
D.~M\"uller, D.~Robaschik, B.~Geyer, F.~M. Dittes and J.~Ho\v{r}ej\v{s}i, \emph{{Wave functions, evolution equations and evolution kernels from light ray operators of QCD}}, \href{https://doi.org/10.1002/prop.2190420202}{\emph{Fortsch. Phys.} {\bfseries 42} (1994) 101--141}, [\href{https://arxiv.org/abs/hep-ph/9812448}{{\ttfamily hep-ph/9812448}}].

\bibitem{Ji:1996ek}
X.-D. Ji, \emph{{Gauge-Invariant Decomposition of Nucleon Spin}}, \href{https://doi.org/10.1103/PhysRevLett.78.610}{\emph{Phys. Rev. Lett.} {\bfseries 78} (1997) 610--613}, [\href{https://arxiv.org/abs/hep-ph/9603249}{{\ttfamily hep-ph/9603249}}].

\bibitem{Ji:1998pc}
X.-D. Ji, \emph{{Off forward parton distributions}}, \href{https://doi.org/10.1088/0954-3899/24/7/002}{\emph{J. Phys. G} {\bfseries 24} (1998) 1181--1205}, [\href{https://arxiv.org/abs/hep-ph/9807358}{{\ttfamily hep-ph/9807358}}].

\bibitem{Burkardt:2000za}
M.~Burkardt, \emph{{Impact parameter dependent parton distributions and off forward parton distributions for zeta ---\ensuremath{>} 0}}, \href{https://doi.org/10.1103/PhysRevD.62.071503}{\emph{Phys. Rev. D} {\bfseries 62} (2000) 071503}, [\href{https://arxiv.org/abs/hep-ph/0005108}{{\ttfamily hep-ph/0005108}}].

\bibitem{Burkardt:2002hr}
M.~Burkardt, \emph{{Impact parameter space interpretation for generalized parton distributions}}, \href{https://doi.org/10.1142/S0217751X03012370}{\emph{Int. J. Mod. Phys. A} {\bfseries 18} (2003) 173--208}, [\href{https://arxiv.org/abs/hep-ph/0207047}{{\ttfamily hep-ph/0207047}}].

\bibitem{Ji:2003ak}
X.-d. Ji, \emph{{Viewing the proton through 'color' filters}}, \href{https://doi.org/10.1103/PhysRevLett.91.062001}{\emph{Phys. Rev. Lett.} {\bfseries 91} (2003) 062001}, [\href{https://arxiv.org/abs/hep-ph/0304037}{{\ttfamily hep-ph/0304037}}].

\bibitem{Belitsky:2003nz}
A.~V. Belitsky, X.-d. Ji and F.~Yuan, \emph{{Quark imaging in the proton via quantum phase space distributions}}, \href{https://doi.org/10.1103/PhysRevD.69.074014}{\emph{Phys. Rev. D} {\bfseries 69} (2004) 074014}, [\href{https://arxiv.org/abs/hep-ph/0307383}{{\ttfamily hep-ph/0307383}}].

\bibitem{Ji:1994av}
X.-D. Ji, \emph{{A QCD analysis of the mass structure of the nucleon}}, \href{https://doi.org/10.1103/PhysRevLett.74.1071}{\emph{Phys. Rev. Lett.} {\bfseries 74} (1995) 1071--1074}, [\href{https://arxiv.org/abs/hep-ph/9410274}{{\ttfamily hep-ph/9410274}}].

\bibitem{Polyakov:2002yz}
M.~V. Polyakov, \emph{{Generalized parton distributions and strong forces inside nucleons and nuclei}}, \href{https://doi.org/10.1016/S0370-2693(03)00036-4}{\emph{Phys. Lett. B} {\bfseries 555} (2003) 57--62}, [\href{https://arxiv.org/abs/hep-ph/0210165}{{\ttfamily hep-ph/0210165}}].

\bibitem{Ji:1996nm}
X.-D. Ji, \emph{{Deeply virtual Compton scattering}}, \href{https://doi.org/10.1103/PhysRevD.55.7114}{\emph{Phys. Rev. D} {\bfseries 55} (1997) 7114--7125}, [\href{https://arxiv.org/abs/hep-ph/9609381}{{\ttfamily hep-ph/9609381}}].

\bibitem{Radyushkin:1996ru}
A.~V. Radyushkin, \emph{{Asymmetric gluon distributions and hard diffractive electroproduction}}, \href{https://doi.org/10.1016/0370-2693(96)00844-1}{\emph{Phys. Lett. B} {\bfseries 385} (1996) 333--342}, [\href{https://arxiv.org/abs/hep-ph/9605431}{{\ttfamily hep-ph/9605431}}].

\bibitem{Collins:1996fb}
J.~C. Collins, L.~Frankfurt and M.~Strikman, \emph{{Factorization for hard exclusive electroproduction of mesons in QCD}}, \href{https://doi.org/10.1103/PhysRevD.56.2982}{\emph{Phys. Rev. D} {\bfseries 56} (1997) 2982--3006}, [\href{https://arxiv.org/abs/hep-ph/9611433}{{\ttfamily hep-ph/9611433}}].

\bibitem{Bertone:2021yyz}
V.~Bertone, H.~Dutrieux, C.~Mezrag, H.~Moutarde and P.~Sznajder, \emph{{Deconvolution problem of deeply virtual Compton scattering}}, \href{https://doi.org/10.1103/PhysRevD.103.114019}{\emph{Phys. Rev. D} {\bfseries 103} (2021) 114019}, [\href{https://arxiv.org/abs/2104.03836}{{\ttfamily 2104.03836}}].

\bibitem{CLAS:2021lky}
{\scshape CLAS} collaboration, P.~Chatagnon et~al., \emph{{First Measurement of Timelike Compton Scattering}}, \href{https://doi.org/10.1103/PhysRevLett.127.262501}{\emph{Phys. Rev. Lett.} {\bfseries 127} (2021) 262501}, [\href{https://arxiv.org/abs/2108.11746}{{\ttfamily 2108.11746}}].

\bibitem{GlueX:2019mkq}
{\scshape GlueX} collaboration, A.~Ali et~al., \emph{{First Measurement of Near-Threshold J/\ensuremath{\psi} Exclusive Photoproduction off the Proton}}, \href{https://doi.org/10.1103/PhysRevLett.123.072001}{\emph{Phys. Rev. Lett.} {\bfseries 123} (2019) 072001}, [\href{https://arxiv.org/abs/1905.10811}{{\ttfamily 1905.10811}}].

\bibitem{Duran:2022xag}
B.~Duran et~al., \emph{{When Color meets Gravity; Near-Threshold Exclusive $J/\psi$ Photoproduction on the Proton}},  \href{https://arxiv.org/abs/2207.05212}{{\ttfamily 2207.05212}}.

\bibitem{GlueX:2023pev}
{\scshape GlueX} collaboration, S.~Adhikari et~al., \emph{{Measurement of the J/$\psi $ photoproduction cross section over the full near-threshold kinematic region}}, \href{https://doi.org/10.1103/PhysRevC.108.025201}{\emph{Phys. Rev. C} {\bfseries 108} (2023) 025201}, [\href{https://arxiv.org/abs/2304.03845}{{\ttfamily 2304.03845}}].

\bibitem{Berger:2001xd}
E.~R. Berger, M.~Diehl and B.~Pire, \emph{{Time - like Compton scattering: Exclusive photoproduction of lepton pairs}}, \href{https://doi.org/10.1007/s100520200917}{\emph{Eur. Phys. J. C} {\bfseries 23} (2002) 675--689}, [\href{https://arxiv.org/abs/hep-ph/0110062}{{\ttfamily hep-ph/0110062}}].

\bibitem{Ivanov:2004vd}
D.~Y. Ivanov, A.~Schafer, L.~Szymanowski and G.~Krasnikov, \emph{{Exclusive photoproduction of a heavy vector meson in QCD}}, \href{https://doi.org/10.1140/epjc/s2004-01712-x}{\emph{Eur. Phys. J. C} {\bfseries 34} (2004) 297--316}, [\href{https://arxiv.org/abs/hep-ph/0401131}{{\ttfamily hep-ph/0401131}}].

\bibitem{Guo:2021ibg}
Y.~Guo, X.~Ji and Y.~Liu, \emph{{QCD Analysis of Near-Threshold Photon-Proton Production of Heavy Quarkonium}}, \href{https://doi.org/10.1103/PhysRevD.103.096010}{\emph{Phys. Rev. D} {\bfseries 103} (2021) 096010}, [\href{https://arxiv.org/abs/2103.11506}{{\ttfamily 2103.11506}}].

\bibitem{Guo:2023pqw}
Y.~Guo, X.~Ji, Y.~Liu and J.~Yang, \emph{{Updated analysis of near-threshold heavy quarkonium production for probe of proton\textquoteright{}s gluonic gravitational form factors}}, \href{https://doi.org/10.1103/PhysRevD.108.034003}{\emph{Phys. Rev. D} {\bfseries 108} (2023) 034003}, [\href{https://arxiv.org/abs/2305.06992}{{\ttfamily 2305.06992}}].

\bibitem{Guo:2023qgu}
Y.~Guo, X.~Ji and F.~Yuan, \emph{{Proton\textquoteright{}s gluon GPDs at large skewness and gravitational form factors from near threshold heavy quarkonium photoproduction}}, \href{https://doi.org/10.1103/PhysRevD.109.014014}{\emph{Phys. Rev. D} {\bfseries 109} (2024) 014014}, [\href{https://arxiv.org/abs/2308.13006}{{\ttfamily 2308.13006}}].

\bibitem{Belitsky:2002tf}
A.~V. Belitsky and D.~Mueller, \emph{{Exclusive electroproduction of lepton pairs as a probe of nucleon structure}}, \href{https://doi.org/10.1103/PhysRevLett.90.022001}{\emph{Phys. Rev. Lett.} {\bfseries 90} (2003) 022001}, [\href{https://arxiv.org/abs/hep-ph/0210313}{{\ttfamily hep-ph/0210313}}].

\bibitem{Guidal:2002kt}
M.~Guidal and M.~Vanderhaeghen, \emph{{Double deeply virtual Compton scattering off the nucleon}}, \href{https://doi.org/10.1103/PhysRevLett.90.012001}{\emph{Phys. Rev. Lett.} {\bfseries 90} (2003) 012001}, [\href{https://arxiv.org/abs/hep-ph/0208275}{{\ttfamily hep-ph/0208275}}].

\bibitem{Pedrak:2017cpp}
A.~Pedrak, B.~Pire, L.~Szymanowski and J.~Wagner, \emph{{Hard photoproduction of a diphoton with a large invariant mass}}, \href{https://doi.org/10.1103/PhysRevD.96.074008}{\emph{Phys. Rev. D} {\bfseries 96} (2017) 074008}, [\href{https://arxiv.org/abs/1708.01043}{{\ttfamily 1708.01043}}].

\bibitem{Duplancic:2023kwe}
G.~Duplan\v{c}i\'c, S.~Nabeebaccus, K.~Passek-Kumeri\v{c}ki, B.~Pire, L.~Szymanowski and S.~Wallon, \emph{{Probing chiral-even and chiral-odd leading twist quark generalized parton distributions through the exclusive photoproduction of a \ensuremath{\gamma}\ensuremath{\rho} pair}}, \href{https://doi.org/10.1103/PhysRevD.107.094023}{\emph{Phys. Rev. D} {\bfseries 107} (2023) 094023}, [\href{https://arxiv.org/abs/2302.12026}{{\ttfamily 2302.12026}}].

\bibitem{Qiu:2022bpq}
J.-W. Qiu and Z.~Yu, \emph{{Exclusive production of a pair of high transverse momentum photons in pion-nucleon collisions for extracting generalized parton distributions}}, \href{https://doi.org/10.1007/JHEP08(2022)103}{\emph{JHEP} {\bfseries 08} (2022) 103}, [\href{https://arxiv.org/abs/2205.07846}{{\ttfamily 2205.07846}}].

\bibitem{Qiu:2022pla}
J.-W. Qiu and Z.~Yu, \emph{{Single diffractive hard exclusive processes for the study of generalized parton distributions}}, \href{https://doi.org/10.1103/PhysRevD.107.014007}{\emph{Phys. Rev. D} {\bfseries 107} (2023) 014007}, [\href{https://arxiv.org/abs/2210.07995}{{\ttfamily 2210.07995}}].

\bibitem{Qiu:2023mrm}
J.-W. Qiu and Z.~Yu, \emph{{Extraction of the Parton Momentum-Fraction Dependence of Generalized Parton Distributions from Exclusive Photoproduction}}, \href{https://doi.org/10.1103/PhysRevLett.131.161902}{\emph{Phys. Rev. Lett.} {\bfseries 131} (2023) 161902}, [\href{https://arxiv.org/abs/2305.15397}{{\ttfamily 2305.15397}}].

\bibitem{Nabeebaccus:2023rzr}
S.~Nabeebaccus, J.~Schoenleber, L.~Szymanowski and S.~Wallon, \emph{{Breakdown of collinear factorization in the exclusive photoproduction of a $ \pi ^{0}\gamma $ pair with large invariant mass}},  \href{https://arxiv.org/abs/2311.09146}{{\ttfamily 2311.09146}}.

\bibitem{Qiu:2024mny}
J.-W. Qiu and Z.~Yu, \emph{{Extracting transition generalized parton distributions from hard exclusive pion-nucleon scattering}}, \href{https://doi.org/10.1103/PhysRevD.109.074023}{\emph{Phys. Rev. D} {\bfseries 109} (2024) 074023}, [\href{https://arxiv.org/abs/2401.13207}{{\ttfamily 2401.13207}}].

\bibitem{AbdulKhalek:2021gbh}
R.~Abdul~Khalek et~al., \emph{{Science Requirements and Detector Concepts for the Electron-Ion Collider: EIC Yellow Report}},  \href{https://arxiv.org/abs/2103.05419}{{\ttfamily 2103.05419}}.

\bibitem{Alexandrou:2021jok}
C.~Alexandrou, S.~Bacchio, M.~Constantinou, J.~Finkenrath, K.~Hadjiyiannakou, K.~Jansen et~al., \emph{{Nucleon form factors from $N_f$=2+1+1 twisted mass QCD at the physical point}}, \href{https://doi.org/10.22323/1.396.0250}{\emph{PoS} {\bfseries LATTICE2021} (2022) 250}, [\href{https://arxiv.org/abs/2112.06750}{{\ttfamily 2112.06750}}].

\bibitem{Hasan:2017wwt}
N.~Hasan, J.~Green, S.~Meinel, M.~Engelhardt, S.~Krieg, J.~Negele et~al., \emph{{Computing the nucleon charge and axial radii directly at $Q^2=0$ in lattice QCD}}, \href{https://doi.org/10.1103/PhysRevD.97.034504}{\emph{Phys. Rev. D} {\bfseries 97} (2018) 034504}, [\href{https://arxiv.org/abs/1711.11385}{{\ttfamily 1711.11385}}].

\bibitem{Shintani:2018ozy}
E.~Shintani, K.-I. Ishikawa, Y.~Kuramashi, S.~Sasaki and T.~Yamazaki, \emph{{Nucleon form factors and root-mean-square radii on a (10.8 fm)$^4$ lattice at the physical point}}, \href{https://doi.org/10.1103/PhysRevD.99.014510}{\emph{Phys. Rev. D} {\bfseries 99} (2019) 014510}, [\href{https://arxiv.org/abs/1811.07292}{{\ttfamily 1811.07292}}].

\bibitem{Jang:2018djx}
{\scshape PNDME} collaboration, Y.-C. Jang, T.~Bhattacharya, R.~Gupta, H.-W. Lin and B.~Yoon, \emph{{Updates on Nucleon Form Factors from Clover-on-HISQ Lattice Formulation}}, \href{https://doi.org/10.22323/1.334.0123}{\emph{PoS} {\bfseries LATTICE2018} (2018) 123}, [\href{https://arxiv.org/abs/1901.00060}{{\ttfamily 1901.00060}}].

\bibitem{Bhattacharya:2023ays}
S.~Bhattacharya, K.~Cichy, M.~Constantinou, X.~Gao, A.~Metz, J.~Miller et~al., \emph{{Moments of proton GPDs from the OPE of nonlocal quark bilinears up to NNLO}}, \href{https://doi.org/10.1103/PhysRevD.108.014507}{\emph{Phys. Rev. D} {\bfseries 108} (2023) 014507}, [\href{https://arxiv.org/abs/2305.11117}{{\ttfamily 2305.11117}}].

\bibitem{Bhattacharya:2023jsc}
S.~Bhattacharya et~al., \emph{{Generalized parton distributions from lattice QCD with asymmetric momentum transfer: Axial-vector case}}, \href{https://doi.org/10.1103/PhysRevD.109.034508}{\emph{Phys. Rev. D} {\bfseries 109} (2024) 034508}, [\href{https://arxiv.org/abs/2310.13114}{{\ttfamily 2310.13114}}].

\bibitem{Pefkou:2021fni}
D.~A. Pefkou, D.~C. Hackett and P.~E. Shanahan, \emph{{Gluon gravitational structure of hadrons of different spin}}, \href{https://doi.org/10.1103/PhysRevD.105.054509}{\emph{Phys. Rev. D} {\bfseries 105} (2022) 054509}, [\href{https://arxiv.org/abs/2107.10368}{{\ttfamily 2107.10368}}].

\bibitem{Hackett:2023rif}
D.~C. Hackett, D.~A. Pefkou and P.~E. Shanahan, \emph{{Gravitational form factors of the proton from lattice QCD}},  \href{https://arxiv.org/abs/2310.08484}{{\ttfamily 2310.08484}}.

\bibitem{Ji:2013dva}
X.~Ji, \emph{{Parton Physics on a Euclidean Lattice}}, \href{https://doi.org/10.1103/PhysRevLett.110.262002}{\emph{Phys. Rev. Lett.} {\bfseries 110} (2013) 262002}, [\href{https://arxiv.org/abs/1305.1539}{{\ttfamily 1305.1539}}].

\bibitem{Ji:2020ect}
X.~Ji, Y.-S. Liu, Y.~Liu, J.-H. Zhang and Y.~Zhao, \emph{{Large-momentum effective theory}}, \href{https://doi.org/10.1103/RevModPhys.93.035005}{\emph{Rev. Mod. Phys.} {\bfseries 93} (2021) 035005}, [\href{https://arxiv.org/abs/2004.03543}{{\ttfamily 2004.03543}}].

\bibitem{Alexandrou:2020zbe}
C.~Alexandrou, K.~Cichy, M.~Constantinou, K.~Hadjiyiannakou, K.~Jansen, A.~Scapellato et~al., \emph{{Unpolarized and helicity generalized parton distributions of the proton within lattice QCD}}, \href{https://doi.org/10.1103/PhysRevLett.125.262001}{\emph{Phys. Rev. Lett.} {\bfseries 125} (2020) 262001}, [\href{https://arxiv.org/abs/2008.10573}{{\ttfamily 2008.10573}}].

\bibitem{Lin:2021brq}
H.-W. Lin, \emph{{Nucleon helicity generalized parton distribution at physical pion mass from lattice QCD}}, \href{https://doi.org/10.1016/j.physletb.2021.136821}{\emph{Phys. Lett. B} {\bfseries 824} (2022) 136821}, [\href{https://arxiv.org/abs/2112.07519}{{\ttfamily 2112.07519}}].

\bibitem{Bhattacharya:2022aob}
S.~Bhattacharya, K.~Cichy, M.~Constantinou, J.~Dodson, X.~Gao, A.~Metz et~al., \emph{{Generalized parton distributions from lattice QCD with asymmetric momentum transfer: Unpolarized quarks}}, \href{https://doi.org/10.1103/PhysRevD.106.114512}{\emph{Phys. Rev. D} {\bfseries 106} (2022) 114512}, [\href{https://arxiv.org/abs/2209.05373}{{\ttfamily 2209.05373}}].

\bibitem{Guo:2022upw}
Y.~Guo, X.~Ji and K.~Shiells, \emph{{Generalized parton distributions through universal moment parameterization: zero skewness case}}, \href{https://doi.org/10.1007/JHEP09(2022)215}{\emph{JHEP} {\bfseries 09} (2022) 215}, [\href{https://arxiv.org/abs/2207.05768}{{\ttfamily 2207.05768}}].

\bibitem{Guo:2023ahv}
Y.~Guo, X.~Ji, M.~G. Santiago, K.~Shiells and J.~Yang, \emph{{Generalized parton distributions through universal moment parameterization: non-zero skewness case}}, \href{https://doi.org/10.1007/JHEP05(2023)150}{\emph{JHEP} {\bfseries 05} (2023) 150}, [\href{https://arxiv.org/abs/2302.07279}{{\ttfamily 2302.07279}}].

\bibitem{Mueller:2005ed}
D.~Mueller and A.~Schafer, \emph{{Complex conformal spin partial wave expansion of generalized parton distributions and distribution amplitudes}}, \href{https://doi.org/10.1016/j.nuclphysb.2006.01.019}{\emph{Nucl. Phys. B} {\bfseries 739} (2006) 1--59}, [\href{https://arxiv.org/abs/hep-ph/0509204}{{\ttfamily hep-ph/0509204}}].

\bibitem{Kumericki:2009uq}
K.~Kumeri\v{c}ki and D.~Mueller, \emph{{Deeply virtual Compton scattering at small $x_B$ and the access to the GPD H}}, \href{https://doi.org/10.1016/j.nuclphysb.2010.07.015}{\emph{Nucl. Phys. B} {\bfseries 841} (2010) 1--58}, [\href{https://arxiv.org/abs/0904.0458}{{\ttfamily 0904.0458}}].

\bibitem{Cocuzza:2022jye}
{\scshape Jefferson Lab Angular Momentum (JAM)} collaboration, C.~Cocuzza, W.~Melnitchouk, A.~Metz and N.~Sato, \emph{{Polarized antimatter in the proton from a global QCD analysis}}, \href{https://doi.org/10.1103/PhysRevD.106.L031502}{\emph{Phys. Rev. D} {\bfseries 106} (2022) L031502}, [\href{https://arxiv.org/abs/2202.03372}{{\ttfamily 2202.03372}}].

\bibitem{Ye:2017gyb}
Z.~Ye, J.~Arrington, R.~J. Hill and G.~Lee, \emph{{Proton and Neutron Electromagnetic Form Factors and Uncertainties}}, \href{https://doi.org/10.1016/j.physletb.2017.11.023}{\emph{Phys. Lett. B} {\bfseries 777} (2018) 8--15}, [\href{https://arxiv.org/abs/1707.09063}{{\ttfamily 1707.09063}}].

\bibitem{CLAS:2018ddh}
{\scshape CLAS} collaboration, M.~Hattawy et~al., \emph{{Exploring the Structure of the Bound Proton with Deeply Virtual Compton Scattering}}, \href{https://doi.org/10.1103/PhysRevLett.123.032502}{\emph{Phys. Rev. Lett.} {\bfseries 123} (2019) 032502}, [\href{https://arxiv.org/abs/1812.07628}{{\ttfamily 1812.07628}}].

\bibitem{CLAS:2021gwi}
{\scshape CLAS} collaboration, V.~Burkert et~al., \emph{{Beam charge asymmetries for deeply virtual Compton scattering off the proton}}, \href{https://doi.org/10.1140/epja/s10050-021-00474-z}{\emph{Eur. Phys. J. A} {\bfseries 57} (2021) 186}, [\href{https://arxiv.org/abs/2103.12651}{{\ttfamily 2103.12651}}].

\bibitem{Georges:2017xjy}
{\scshape Jefferson Lab Hall A} collaboration, F.~Georges, \emph{{Deeply Virtual Compton Scattering at 11GeV in Jefferson Lab Hall A}}, \href{https://doi.org/10.22323/1.310.0170}{\emph{PoS} {\bfseries Hadron2017} (2018) 170}.

\bibitem{JeffersonLabHallA:2022pnx}
{\scshape Jefferson Lab Hall A} collaboration, F.~Georges et~al., \emph{{Deeply Virtual Compton Scattering Cross Section at High Bjorken xB}}, \href{https://doi.org/10.1103/PhysRevLett.128.252002}{\emph{Phys. Rev. Lett.} {\bfseries 128} (2022) 252002}, [\href{https://arxiv.org/abs/2201.03714}{{\ttfamily 2201.03714}}].

\bibitem{H1:2009wnw}
{\scshape H1} collaboration, F.~D. Aaron et~al., \emph{{Deeply Virtual Compton Scattering and its Beam Charge Asymmetry in e+- Collisions at HERA}}, \href{https://doi.org/10.1016/j.physletb.2009.10.035}{\emph{Phys. Lett. B} {\bfseries 681} (2009) 391--399}, [\href{https://arxiv.org/abs/0907.5289}{{\ttfamily 0907.5289}}].

\bibitem{H1:2005dtp}
{\scshape H1} collaboration, A.~Aktas et~al., \emph{{Elastic J/psi production at HERA}}, \href{https://doi.org/10.1140/epjc/s2006-02519-5}{\emph{Eur. Phys. J. C} {\bfseries 46} (2006) 585--603}, [\href{https://arxiv.org/abs/hep-ex/0510016}{{\ttfamily hep-ex/0510016}}].

\bibitem{Ryskin:1992ui}
M.~G. Ryskin, \emph{{Diffractive J / psi electroproduction in LLA QCD}}, \href{https://doi.org/10.1007/BF01555742}{\emph{Z. Phys. C} {\bfseries 57} (1993) 89--92}.

\bibitem{Brodsky:1994kf}
S.~J. Brodsky, L.~Frankfurt, J.~F. Gunion, A.~H. Mueller and M.~Strikman, \emph{{Diffractive leptoproduction of vector mesons in QCD}}, \href{https://doi.org/10.1103/PhysRevD.50.3134}{\emph{Phys. Rev. D} {\bfseries 50} (1994) 3134--3144}, [\href{https://arxiv.org/abs/hep-ph/9402283}{{\ttfamily hep-ph/9402283}}].

\bibitem{Frankfurt:1997fj}
L.~Frankfurt, W.~Koepf and M.~Strikman, \emph{{Diffractive heavy quarkonium photoproduction and electroproduction in QCD}}, \href{https://doi.org/10.1103/PhysRevD.57.512}{\emph{Phys. Rev. D} {\bfseries 57} (1998) 512--526}, [\href{https://arxiv.org/abs/hep-ph/9702216}{{\ttfamily hep-ph/9702216}}].

\bibitem{Frankfurt:2000ez}
L.~Frankfurt, M.~McDermott and M.~Strikman, \emph{{A Fresh look at diffractive J / psi photoproduction at HERA, with predictions for THERA}}, \href{https://doi.org/10.1088/1126-6708/2001/03/045}{\emph{JHEP} {\bfseries 03} (2001) 045}, [\href{https://arxiv.org/abs/hep-ph/0009086}{{\ttfamily hep-ph/0009086}}].

\bibitem{Kowalski:2006hc}
H.~Kowalski, L.~Motyka and G.~Watt, \emph{{Exclusive diffractive processes at HERA within the dipole picture}}, \href{https://doi.org/10.1103/PhysRevD.74.074016}{\emph{Phys. Rev. D} {\bfseries 74} (2006) 074016}, [\href{https://arxiv.org/abs/hep-ph/0606272}{{\ttfamily hep-ph/0606272}}].

\bibitem{Chen:2019uit}
Z.-Q. Chen and C.-F. Qiao, \emph{{NLO QCD corrections to exclusive electroproduction of quarkonium}}, \href{https://doi.org/10.1016/j.physletb.2019.134816}{\emph{Phys. Lett. B} {\bfseries 797} (2019) 134816}, [\href{https://arxiv.org/abs/1903.00171}{{\ttfamily 1903.00171}}].

\bibitem{Koempel:2011rc}
J.~Koempel, P.~Kroll, A.~Metz and J.~Zhou, \emph{{Exclusive production of quarkonia as a probe of the GPD E for gluons}}, \href{https://doi.org/10.1103/PhysRevD.85.051502}{\emph{Phys. Rev. D} {\bfseries 85} (2012) 051502}, [\href{https://arxiv.org/abs/1112.1334}{{\ttfamily 1112.1334}}].

\bibitem{Koempel:2015xol}
J.~P. Koempel, \emph{{Exclusive Production of Quarkonia and Generalized Parton Distributions}}, Ph.D. thesis, Temple U., 2015.

\bibitem{Flett:2021ghh}
C.~A. Flett, J.~A. Gracey, S.~P. Jones and T.~Teubner, \emph{{Exclusive heavy vector meson electroproduction to NLO in collinear factorisation}}, \href{https://doi.org/10.1007/JHEP08(2021)150}{\emph{JHEP} {\bfseries 08} (2021) 150}, [\href{https://arxiv.org/abs/2105.07657}{{\ttfamily 2105.07657}}].

\bibitem{Mantysaari:2021ryb}
H.~M\"antysaari and J.~Penttala, \emph{{Exclusive heavy vector meson production at next-to-leading order in the dipole picture}}, \href{https://doi.org/10.1016/j.physletb.2021.136723}{\emph{Phys. Lett. B} {\bfseries 823} (2021) 136723}, [\href{https://arxiv.org/abs/2104.02349}{{\ttfamily 2104.02349}}].

\bibitem{Mantysaari:2022kdm}
H.~M\"antysaari and J.~Penttala, \emph{{Complete calculation of exclusive heavy vector meson production at next-to-leading order in the dipole picture}}, \href{https://doi.org/10.1007/JHEP08(2022)247}{\emph{JHEP} {\bfseries 08} (2022) 247}, [\href{https://arxiv.org/abs/2204.14031}{{\ttfamily 2204.14031}}].

\bibitem{Eskola:2022vpi}
K.~J. Eskola, C.~A. Flett, V.~Guzey, T.~L\"oyt\"ainen and H.~Paukkunen, \emph{{Exclusive J/\ensuremath{\psi} photoproduction in ultraperipheral Pb+Pb collisions at the CERN Large Hadron Collider calculated at next-to-leading order perturbative QCD}}, \href{https://doi.org/10.1103/PhysRevC.106.035202}{\emph{Phys. Rev. C} {\bfseries 106} (2022) 035202}, [\href{https://arxiv.org/abs/2203.11613}{{\ttfamily 2203.11613}}].

\bibitem{Goloskokov:2024egn}
S.~V. Goloskokov, Y.-P. Xie and X.~Chen, \emph{{Study of gluon GPDs in exclusive $J/\psi$ production in electron-proton scattering}},  \href{https://arxiv.org/abs/2408.05800}{{\ttfamily 2408.05800}}.

\bibitem{Flett:2024htj}
C.~A. Flett, J.~P. Lansberg, S.~Nabeebaccus, M.~Nefedov, P.~Sznajder and J.~Wagner, \emph{{Exclusive vector-quarkonium photoproduction at NLO in alpha\_s in collinear factorisation with evolution of the generalised parton distributions and high-energy resummation}},  \href{https://arxiv.org/abs/2409.05738}{{\ttfamily 2409.05738}}.

\bibitem{Goloskokov:2006hr}
S.~V. Goloskokov and P.~Kroll, \emph{{The Longitudinal cross-section of vector meson electroproduction}}, \href{https://doi.org/10.1140/epjc/s10052-007-0228-4}{\emph{Eur. Phys. J. C} {\bfseries 50} (2007) 829--842}, [\href{https://arxiv.org/abs/hep-ph/0611290}{{\ttfamily hep-ph/0611290}}].

\bibitem{Meskauskas:2011aa}
M.~Meskauskas and D.~M\"uller, \emph{{A Fresh Look at Exclusive Electroproduction of Light Vector Mesons}}, \href{https://doi.org/10.1140/epjc/s10052-014-2719-4}{\emph{Eur. Phys. J. C} {\bfseries 74} (2014) 2719}, [\href{https://arxiv.org/abs/1112.2597}{{\ttfamily 1112.2597}}].

\bibitem{Lautenschlager:2013uya}
T.~Lautenschlager, D.~Muller and A.~Schaefer, \emph{{Global analysis of generalized parton distributions -- collider kinematics --}},  \href{https://arxiv.org/abs/1312.5493}{{\ttfamily 1312.5493}}.

\bibitem{Muller:2013jur}
D.~M\"uller, T.~Lautenschlager, K.~Passek-Kumericki and A.~Schaefer, \emph{{Towards a fitting procedure to deeply virtual meson production - the next-to-leading order case}}, \href{https://doi.org/10.1016/j.nuclphysb.2014.04.012}{\emph{Nucl. Phys. B} {\bfseries 884} (2014) 438--546}, [\href{https://arxiv.org/abs/1310.5394}{{\ttfamily 1310.5394}}].

\bibitem{Duplancic:2016bge}
G.~Duplan\v{c}i\'c, D.~M\"uller and K.~Passek-Kumeri\v{c}ki, \emph{{Next-to-leading order corrections to deeply virtual production of pseudoscalar mesons}}, \href{https://doi.org/10.1016/j.physletb.2017.05.097}{\emph{Phys. Lett. B} {\bfseries 771} (2017) 603--610}, [\href{https://arxiv.org/abs/1612.01937}{{\ttfamily 1612.01937}}].

\bibitem{Cuic:2023mki}
M.~\v{C}ui\'c, G.~Duplan\v{c}i\'c, K.~Kumeri\v{c}ki and K.~Passek-K., \emph{{NLO corrections to the deeply virtual meson production revisited: impact on the extraction of generalized parton distributions}},  \href{https://arxiv.org/abs/2310.13837}{{\ttfamily 2310.13837}}.

\bibitem{Kovchegov:1999ji}
Y.~V. Kovchegov and E.~Levin, \emph{{Diffractive dissociation including multiple pomeron exchanges in high parton density QCD}}, \href{https://doi.org/10.1016/S0550-3213(00)00125-5}{\emph{Nucl. Phys. B} {\bfseries 577} (2000) 221--239}, [\href{https://arxiv.org/abs/hep-ph/9911523}{{\ttfamily hep-ph/9911523}}].

\bibitem{Kovner:2001vi}
A.~Kovner and U.~A. Wiedemann, \emph{{Eikonal evolution and gluon radiation}}, \href{https://doi.org/10.1103/PhysRevD.64.114002}{\emph{Phys. Rev. D} {\bfseries 64} (2001) 114002}, [\href{https://arxiv.org/abs/hep-ph/0106240}{{\ttfamily hep-ph/0106240}}].

\bibitem{Hentschinski:2005er}
M.~Hentschinski, H.~Weigert and A.~Schafer, \emph{{Extension of the color glass condensate approach to diffractive reactions}}, \href{https://doi.org/10.1103/PhysRevD.73.051501}{\emph{Phys. Rev. D} {\bfseries 73} (2006) 051501}, [\href{https://arxiv.org/abs/hep-ph/0509272}{{\ttfamily hep-ph/0509272}}].

\bibitem{Kovner:2006ge}
A.~Kovner, M.~Lublinsky and H.~Weigert, \emph{{Treading on the cut: Semi inclusive observables at high energy}}, \href{https://doi.org/10.1103/PhysRevD.74.114023}{\emph{Phys. Rev. D} {\bfseries 74} (2006) 114023}, [\href{https://arxiv.org/abs/hep-ph/0608258}{{\ttfamily hep-ph/0608258}}].

\bibitem{Hatta:2006hs}
Y.~Hatta, E.~Iancu, C.~Marquet, G.~Soyez and D.~N. Triantafyllopoulos, \emph{{Diffusive scaling and the high-energy limit of deep inelastic scattering in QCD at large N(c)}}, \href{https://doi.org/10.1016/j.nuclphysa.2006.04.003}{\emph{Nucl. Phys. A} {\bfseries 773} (2006) 95--155}, [\href{https://arxiv.org/abs/hep-ph/0601150}{{\ttfamily hep-ph/0601150}}].

\bibitem{Rezaeian:2012ji}
A.~H. Rezaeian, M.~Siddikov, M.~Van~de Klundert and R.~Venugopalan, \emph{{Analysis of combined HERA data in the Impact-Parameter dependent Saturation model}}, \href{https://doi.org/10.1103/PhysRevD.87.034002}{\emph{Phys. Rev. D} {\bfseries 87} (2013) 034002}, [\href{https://arxiv.org/abs/1212.2974}{{\ttfamily 1212.2974}}].

\bibitem{Hatta:2017cte}
Y.~Hatta, B.-W. Xiao and F.~Yuan, \emph{{Gluon Tomography from Deeply Virtual Compton Scattering at Small-x}}, \href{https://doi.org/10.1103/PhysRevD.95.114026}{\emph{Phys. Rev. D} {\bfseries 95} (2017) 114026}, [\href{https://arxiv.org/abs/1703.02085}{{\ttfamily 1703.02085}}].

\bibitem{Frankfurt:1999fp}
L.~L. Frankfurt, P.~V. Pobylitsa, M.~V. Polyakov and M.~Strikman, \emph{{Hard exclusive pseudoscalar meson electroproduction and spin structure of a nucleon}}, \href{https://doi.org/10.1103/PhysRevD.60.014010}{\emph{Phys. Rev. D} {\bfseries 60} (1999) 014010}, [\href{https://arxiv.org/abs/hep-ph/9901429}{{\ttfamily hep-ph/9901429}}].

\bibitem{Kroll:2019wug}
P.~Kroll, \emph{{Hard exclusive processes involving kaons}}, \href{https://doi.org/10.1140/epja/i2019-12747-9}{\emph{Eur. Phys. J. A} {\bfseries 55} (2019) 76}, [\href{https://arxiv.org/abs/1901.11380}{{\ttfamily 1901.11380}}].

\bibitem{Ball:2006eu}
P.~Ball, G.~W. Jones and R.~Zwicky, \emph{{$B \to V \gamma$ beyond QCD factorisation}}, \href{https://doi.org/10.1103/PhysRevD.75.054004}{\emph{Phys. Rev. D} {\bfseries 75} (2007) 054004}, [\href{https://arxiv.org/abs/hep-ph/0612081}{{\ttfamily hep-ph/0612081}}].

\bibitem{ParticleDataGroup:2020ssz}
{\scshape Particle Data Group} collaboration, P.~A. Zyla et~al., \emph{{Review of Particle Physics}}, \href{https://doi.org/10.1093/ptep/ptaa104}{\emph{PTEP} {\bfseries 2020} (2020) 083C01}.

\bibitem{Ball:1998sk}
P.~Ball, V.~M. Braun, Y.~Koike and K.~Tanaka, \emph{{Higher twist distribution amplitudes of vector mesons in QCD: Formalism and twist - three distributions}}, \href{https://doi.org/10.1016/S0550-3213(98)00356-3}{\emph{Nucl. Phys. B} {\bfseries 529} (1998) 323--382}, [\href{https://arxiv.org/abs/hep-ph/9802299}{{\ttfamily hep-ph/9802299}}].

\bibitem{Ball:1998ff}
P.~Ball and V.~M. Braun, \emph{{Higher twist distribution amplitudes of vector mesons in QCD: Twist - 4 distributions and meson mass corrections}}, \href{https://doi.org/10.1016/S0550-3213(99)00014-0}{\emph{Nucl. Phys. B} {\bfseries 543} (1999) 201--238}, [\href{https://arxiv.org/abs/hep-ph/9810475}{{\ttfamily hep-ph/9810475}}].

\bibitem{Diehl:2003ny}
M.~Diehl, \emph{{Generalized parton distributions}}, \href{https://doi.org/10.1016/j.physrep.2003.08.002}{\emph{Phys. Rept.} {\bfseries 388} (2003) 41--277}, [\href{https://arxiv.org/abs/hep-ph/0307382}{{\ttfamily hep-ph/0307382}}].

\bibitem{ZEUS:2004yeh}
{\scshape ZEUS} collaboration, S.~Chekanov et~al., \emph{{Exclusive electroproduction of J/psi mesons at HERA}}, \href{https://doi.org/10.1016/j.nuclphysb.2004.06.034}{\emph{Nucl. Phys. B} {\bfseries 695} (2004) 3--37}, [\href{https://arxiv.org/abs/hep-ex/0404008}{{\ttfamily hep-ex/0404008}}].

\bibitem{Bodwin:1994jh}
G.~T. Bodwin, E.~Braaten and G.~P. Lepage, \emph{{Rigorous QCD analysis of inclusive annihilation and production of heavy quarkonium}}, \href{https://doi.org/10.1103/PhysRevD.55.5853}{\emph{Phys. Rev. D} {\bfseries 51} (1995) 1125--1171}, [\href{https://arxiv.org/abs/hep-ph/9407339}{{\ttfamily hep-ph/9407339}}].

\bibitem{Hoodbhoy:1996zg}
P.~Hoodbhoy, \emph{{Wave function corrections and off forward gluon distributions in diffractive J / psi electroproduction}}, \href{https://doi.org/10.1103/PhysRevD.56.388}{\emph{Phys. Rev. D} {\bfseries 56} (1997) 388--393}, [\href{https://arxiv.org/abs/hep-ph/9611207}{{\ttfamily hep-ph/9611207}}].

\bibitem{Kumericki:2007sa}
K.~Kumericki, D.~Mueller and K.~Passek-Kumericki, \emph{{Towards a fitting procedure for deeply virtual Compton scattering at next-to-leading order and beyond}}, \href{https://doi.org/10.1016/j.nuclphysb.2007.10.029}{\emph{Nucl. Phys. B} {\bfseries 794} (2008) 244--323}, [\href{https://arxiv.org/abs/hep-ph/0703179}{{\ttfamily hep-ph/0703179}}].

\bibitem{Cuic:2020iwt}
M.~\v{C}ui\'c, K.~Kumeri\v{c}ki and A.~Sch\"afer, \emph{{Separation of Quark Flavors Using Deeply Virtual Compton Scattering Data}}, \href{https://doi.org/10.1103/PhysRevLett.125.232005}{\emph{Phys. Rev. Lett.} {\bfseries 125} (2020) 232005}, [\href{https://arxiv.org/abs/2007.00029}{{\ttfamily 2007.00029}}].

\bibitem{Han:2024min}
X.-Y. Han, J.~Hua, X.~Ji, C.-D. L\"u, W.~Wang, J.~Xu et~al., \emph{{A new method to access heavy meson lightcone distribution amplitudes from first-principle}},  \href{https://arxiv.org/abs/2403.17492}{{\ttfamily 2403.17492}}.

\bibitem{Curci:1980uw}
G.~Curci, W.~Furmanski and R.~Petronzio, \emph{{Evolution of Parton Densities Beyond Leading Order: The Nonsinglet Case}}, \href{https://doi.org/10.1016/0550-3213(80)90003-6}{\emph{Nucl. Phys. B} {\bfseries 175} (1980) 27--92}.

\bibitem{Belitsky:1998uk}
A.~V. Belitsky, D.~Mueller, L.~Niedermeier and A.~Schafer, \emph{{Evolution of nonforward parton distributions in next-to-leading order: Singlet sector}}, \href{https://doi.org/10.1016/S0550-3213(99)00045-0}{\emph{Nucl. Phys. B} {\bfseries 546} (1999) 279--298}, [\href{https://arxiv.org/abs/hep-ph/9810275}{{\ttfamily hep-ph/9810275}}].

\bibitem{iminuit}
H.~Dembinski and P.~O. et~al., \emph{scikit-hep/iminuit}, .

\bibitem{James:1975dr}
F.~James and M.~Roos, \emph{{Minuit: A System for Function Minimization and Analysis of the Parameter Errors and Correlations}}, \href{https://doi.org/10.1016/0010-4655(75)90039-9}{\emph{Comput. Phys. Commun.} {\bfseries 10} (1975) 343--367}.

\bibitem{Toll:2012mb}
T.~Toll and T.~Ullrich, \emph{{Exclusive diffractive processes in electron-ion collisions}}, \href{https://doi.org/10.1103/PhysRevC.87.024913}{\emph{Phys. Rev. C} {\bfseries 87} (2013) 024913}, [\href{https://arxiv.org/abs/1211.3048}{{\ttfamily 1211.3048}}].

\bibitem{Mantysaari:2016jaz}
H.~M\"antysaari and B.~Schenke, \emph{{Revealing proton shape fluctuations with incoherent diffraction at high energy}}, \href{https://doi.org/10.1103/PhysRevD.94.034042}{\emph{Phys. Rev. D} {\bfseries 94} (2016) 034042}, [\href{https://arxiv.org/abs/1607.01711}{{\ttfamily 1607.01711}}].

\bibitem{gepard}
K.~Kumeri\v{c}ki, ``\textit{Gepard: Tool for studying the 3D quark and gluon distributions in the nucleon}.'' \url{https://gepard.phy.hr}.

\bibitem{Zhang:2024djl}
H.-C. Zhang and X.~Ji, \emph{{On convergence properties of GPD expansion through Mellin/conformal moments and orthogonal polynomials}},  \href{https://arxiv.org/abs/2408.04133}{{\ttfamily 2408.04133}}.

\bibitem{Guo:2022gumpgit}
Y.~Guo et~al., ``{GUMP GPD Global Analysis}.'' \url{https://github.com/yuxunguo/GUMP-Global-GPDs}, 2022.

\end{thebibliography}\endgroup

\end{document}